\documentclass[dvips,12pt,aps,preprint,showpacs]{revtex4}
\usepackage{graphicx}
\input{FEYNMAN.tex}
\begin{document}
                      
\title{Study of the decay $D^0 \to K^- K^- K^+ \pi^+$}
\author{
    E.~M.~Aitala,$^9$
       S.~Amato,$^1$
    J.~C.~Anjos,$^1$
    J.~A.~Appel,$^5$
       D.~Ashery,$^{14}$
       S.~Banerjee,$^5$
       I.~Bediaga,$^1$
       G.~Blaylock,$^8$
    S.~B.~Bracker,$^{15}$
    P.~R.~Burchat,$^{13}$
    R.~A.~Burnstein,$^6$
       T.~Carter,$^5$
 H.~S.~Carvalho,$^{1}$
  N.~K.~Copty,$^{12}$
    L.~M.~Cremaldi,$^9$
 C.~Darling,$^{18}$
       K.~Denisenko,$^5$
       S.~Devmal,$^3$
       A.~Fernandez,$^{11}$
       G.~F.~Fox,$^{12}$
       P.~Gagnon,$^2$
       S.~Gerzon,$^{14}$
       C.~Gobel,$^1$
       K.~Gounder,$^9$
     A.~M.~Halling,$^5$
       G.~Herrera,$^4$
 G.~Hurvits,$^{14}$
       C.~James,$^5$
    P.~A.~Kasper,$^6$
       S.~Kwan,$^5$
    D.~C.~Langs,$^{12}$
       J.~Leslie,$^2$
       J.~Lichtenstadt,$^{14}$
       B.~Lundberg,$^5$
       S.~MayTal-Beck,$^{14}$
       B.~Meadows,$^3$
 J.~R.~T.~de~Mello~Neto,$^1$
    D.~Mihalcea,$^{7}$
    R.~H.~Milburn,$^{16}$
 J.~M.~de~Miranda,$^1$
       A.~Napier,$^{16}$
       A.~Nguyen,$^7$
  A.~B.~d'Oliveira,$^{3,11}$
       K.~O'Shaughnessy,$^2$
    K.~C.~Peng,$^6$
    L.~P.~Perera,$^3$
    M.~V.~Purohit,$^{12}$
       B.~Quinn,$^9$
       S.~Radeztsky,$^{17}$
       A.~Rafatian,$^9$
    N.~W.~Reay,$^7$
    J.~J.~Reidy,$^9$
    A.~C.~dos Reis,$^1$
    H.~A.~Rubin,$^6$
 D.~A.~Sanders,$^9$
 A.~K.~S.~Santha,$^3$
 A.~F.~S.~Santoro,$^1$
       A.~J.~Schwartz,$^{3}$
       M.~Sheaff,$^{17}$
    R.~A.~Sidwell,$^7$
    A.~J.~Slaughter,$^{18}$
    M.~D.~Sokoloff,$^3$
       J.~Solano,$^{1}$
       N.~R.~Stanton,$^7$
      R.~J.~Stefanski,$^5$ 
      K.~Stenson,$^{17}$
    D.~J.~Summers,$^9$
 S.~Takach,$^{18}$
       K.~Thorne,$^5$
    A.~K.~Tripathi,$^{7}$
       S.~Watanabe,$^{17}$
 R.~Weiss-Babai,$^{14}$
       J.~Wiener,$^{10}$
       N.~Witchey,$^7$
       E.~Wolin,$^{18}$
     S.~M.~Yang,$^{7}$
       D.~Yi,$^9$
       S.~Yoshida,$^{7}$
       R.~Zaliznyak,$^{13}$
       and
       C.~Zhang$^7$ \\
(Fermilab E791 Collaboration)
}

\affiliation{
$^1$ Centro Brasileiro de Pesquisas F\'\i sicas, Rio de Janeiro, Brazil\\
$^2$ University of California, Santa Cruz, California 95064\\
$^3$ University of Cincinnati, Cincinnati, Ohio 45221\\
$^4$ CINVESTAV, Mexico City, Mexico\\
$^5$ Fermilab, Batavia, Illinois 60510\\
$^6$ Illinois Institute of Technology, Chicago, Illinois 60616\\
$^7$ Kansas State University, Manhattan, Kansas 66506\\
$^8$ University of Massachusetts, Amherst, Massachusetts 01003\\
$^9$ University of Mississippi-Oxford, University, Mississippi 38677\\
$^{10}$ Princeton University, Princeton, New Jersey 08544\\
$^{11}$ Universidad Autonoma de Puebla, Puebla, Mexico\\
$^{12}$ University of South Carolina, Columbia, South Carolina 29208\\
$^{13}$ Stanford University, Stanford, California 94305\\
$^{14}$ Tel Aviv University, Tel Aviv, Israel\\
$^{15}$ Box 1290, Enderby, BC, VOE 1V0, Canada\\
$^{16}$ Tufts University, Medford, Massachusetts 02155\\
$^{17}$ University of Wisconsin, Madison, Wisconsin 53706\\
$^{18}$ Yale University, New Haven, Connecticut 06511\\
}

\date{\today}
\vspace*{0.2in}
\vskip 0.2in

\begin{abstract}

Using data from the E791 fixed-target hadroproduction
experiment at Fermilab
we have studied the Cabibbo-favored but phase-space suppressed
decay, $ D^0 \to K^- K^- K^+ \pi^+ $.
We find the decay rate  for this mode to
be $( 0.54 \pm  0.16 \pm 0.08) \times 10^{-2} $ times
that for the normalization
mode $ D^0 \to K^- \pi^- \pi^+ \pi^+ $.
We observe a clear signal for $ D^0 \to \phi K^- \pi^+ $
which is consistent with producing $ 0.7 \pm 0.3 $ of the
$ D^0 \to K^- K^- K^+ \pi^+ $ signal.
In the context of simple models, we use our measurements to
estimate the importance of decay amplitudes that
produce extra quark-antiquark pairs from the vacuum relative to
those that do  not.

\end{abstract}

\pacs{13.25.Ft}

\maketitle

The decays $ D^0 \to K^- K^- K^+ \pi^+ $
and $ D^0 \to K^- \pi^- \pi^+ \pi^+ $ are both
Cabibbo-favored,
but phase-space suppresses the former relative to the latter.
In addition, the decay $ D^0 \to K^- K^- K^+ \pi^+ $
requires the production of at least one extra
quark-antiquark pair,
an $ s \overline s $, either from the vacuum or
via a final state interaction.
The more common decay $ D^0 \to K^- \pi^- \pi^+ \pi^+ $
may proceed both via an intermediate state such
as $ {\overline K}^{*0} \rho^0 $ in which the resonant
particles contain only quarks produced directly from
a spectator amplitude, and via an amplitude that requires
the production of at least one extra $ q \overline q $ pair
from the vacuum.

In this paper we present a decay rate measurement 
for $ D^0 \to K^- K^- K^+ \pi^+ $ relative 
to that for $ D^0 \to K^- \pi^- \pi^+ \pi^+ $ using data
from  the E791 fixed-target hadroproduction experiment at
Fermilab.
This allows us to determine the importance of decay amplitudes
that produce extra $ q \overline q $ pairs from the
vacuum relative to those that do not.
In addition, we study the $ K^- K^+ $ invariant mass
distribution in signal events to search 
for intermediate $ \phi $ production.

\leftline{\bf Experimental Overview}

Experiment E791 recorded $2\times10^{10}$
interactions  during the 1991/92 fixed-target run at Fermilab
using a
500 GeV/$c$  $\pi^-$ beam and
an open
geometry spectrometer\cite{e791} in the Tagged Photon Laboratory. 
The target
consisted of one platinum foil and four diamond foils, separated by gaps of 
1.34 to 1.39 cm. Each foil was approximately 0.4\% of a pion interaction length 
thick (0.5 mm for platinum  and 1.6 mm  for carbon). The average decay length
of an 80 GeV $ D^0 $ is approximately 5 mm, so most of the $D^0$'s  decayed
in the air gaps between target foils where backgrounds are lower. Six planes
of silicon microstrip detectors (SMDs) and eight proportional wire  chambers 
(PWCs)  were used to track the beam particles. The downstream detector 
consisted of 17 planes of SMDs for vertex detection, 35 drift chamber planes,
two PWCs,  two magnets 
for momentum analysis (both bending in the same direction),
two multicell threshold 
\v{C}erenkov counters\cite{cerenkov} 
for charged  particle identification
(with nominal pion thresholds of 6 GeV/$c$ and 11 GeV/$c$),
electromagnetic and hadronic calorimeters for electron identification
and 
for  online triggering, and two planes of muon scintillators. 
An interaction trigger  required a beam particle and
an interaction in the target. A very loose  transverse energy trigger, based
on the energy deposited in the calorimeters, and a fast data acquisition 
system \cite{daq} allowed the experiment
to collect data at a rate of 30 Mbytes/s with 
50 $\mu$s/event dead time and to write data to tape at a rate of 10 Mbytes/s.

\bigbreak
\leftline{\bf Event Selection}
Data reconstruction and additional event selection were done using offline
parallel processing systems \cite{farm}. 
Events with evidence of well-separated 
production (primary) and decay (secondary) vertices were retained for further
analysis. 
Candidate $D^0 \to K^- K^- K^+ \pi^+$  and  
$D^0 \to K^-\pi^-\pi^+\pi^+$ decays
(and their corresponding charge conjugate decays, which we include implicitly
whenever we refer to a decay chain)
were selected from events with at least
one candidate four-prong secondary vertex. 
Selection criteria (cuts), used for  both modes, 
were  determined by optimizing the  expected statistical significance  of the 
$ D^0 \to K^- K^- K^+ \pi^+ $ signal.
To avoid bias, we masked the signal region (1.845 GeV/$c^2$  
$ < {\rm mass}(K^- K^- K^+ \pi^+) < $
 1.885 GeV/$c^2$)
in the real data 
early in the analysis and systematically studied sensitivity using a 
combination of real data for background and Monte Carlo simulations
for signal.
Only after we had determined a set of cuts for
the final analysis, and looser and tighter sets of
cuts for studies of systematic uncertainties, did we examine
the data in the signal region.

We used Monte Carlo simulations of $D^0 \to K^- K^- K^+ \pi^+$
decays  and $D^0 \to K^-\pi^-\pi^+\pi^+ $ real data
to estimate the efficiencies of
potential cuts.
We compared Monte Carlo simulations of $D^0 \to K^-\pi^-\pi^+\pi^+$ 
to the real data in this channel to validate our Monte Carlo.
Where the distributions in the Monte Carlo simulation 
of $D^0 \to K^-\pi^-\pi^+\pi^+$
match the real data, we trust the  $D^0 \to K^- K^- K^+ \pi^+$
Monte Carlo simulation.
Where the distributions in the Monte Carlo simulation of  
$D^0 \to K^- K^- K^+ \pi^+$
match those in the Monte Carlo simulation of $D^0 \to K^-\pi^-\pi^+\pi^+ $,
we trust that the corresponding distributions observed for
real $D^0 \to K^-\pi^-\pi^+\pi^+ $ correctly predict those for
$D^0 \to K^- K^- K^+ \pi^+$.
At this stage, we estimated the background within the signal region by interpolating
linearly from sidebands in the 
$ K^- K^- K^+ \pi^+ $ invariant mass distribution
of the data.

\begin{figure}[!]
   \begin{minipage}{2.95in}
    \centering
    \includegraphics[width=2.95in]{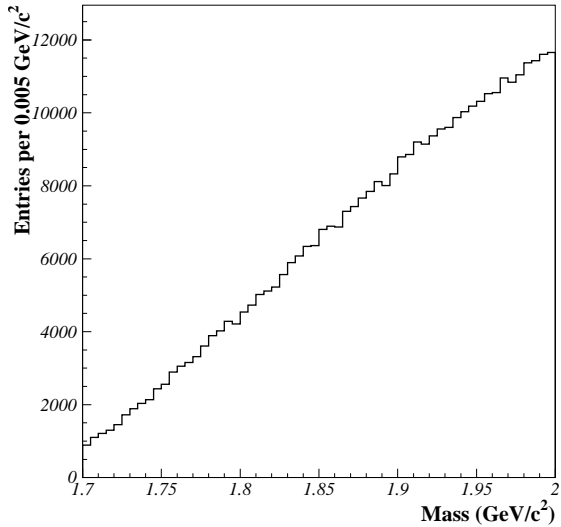}
    \caption{$ K K K \pi $ invariant mass distributions after the
     preliminary event selection}
    \label{microMass}
   \end{minipage}
\end{figure}\medskip

When events were initially reconstructed, a topological vertexing 
algorithm was used to identify 
a primary (interaction) vertex and possible secondary (downstream)
vertices.
Our $ D^0 $ candidates are constructed from four-track
secondary vertices (referred to as `SEED4' candidates) and from three-track
secondary vertices with the addition of a fourth track
(referred to as `SEED3' candidates).
Because the initial topological vertexing algorithm assigned each
track to one vertex candidate, and because it was optimized
for two-body and three-body charm decays, about half of our
signal comes from the SEED4 sample and half from the SEED3 sample.

The signal was expected to be small (between 10 and 50
signal events, depending on cuts and allowing for some uncertainty
in the relative branching ratio),
and the
signal-to-background ratio is better in the SEED4 sample when
all
other cuts are fixed, so the cuts are studied separately 
for SEED4 and SEED3 candidates.
Within each category, we made initial, very loose cuts on a series of
candidate variables, primarily informed by  our prior experience with
similar analyses.
The invariant mass distribution of all candidates
surviving these loose cuts
is shown in Fig.\  \ref{microMass}.
At this point in the analysis we decided to proceed blindly:
to mask the signal region in the data and optimize sensitivity as discussed
in detail below.

As part of the optimization, we create combined figure-of-merit (FOM)
variables from 
kinematic and particle identification variables that are essentially 
independent of each other
in our Monte Carlo simulations and in the $D^0 \to K^-\pi^-\pi^+\pi^+$
data.
For each of the SEED4 and SEED3 samples, 
we start with a set of independent cuts on individual
variables, and then construct combined FOM variables for each of the 
surviving events.
The procedure for constructing FOM variables will be described in
detail after a discussion of the most important variables used in
selecting candidates.

The decay  vertex is  required  to lie outside 
the target foils and other solid material
and to be well-separated from the primary vertex,
with $ \Delta z \ > \  10 \, \sigma_z$ for SEED4 candidates
(where $\sigma_z$ is the error on
$ \Delta z $, the
longitudinal separation between two vertices)
and with  $ \Delta z \ > \  12 \, \sigma_z$ for SEED3 candidates.
The transverse  momentum of the $ D^0 $ candidate with respect to
the line-of-flight defined by the secondary and primary vertex
positions ($p_T$-balance) is required to be less than 250 MeV/$c$.
The transverse distance of closest approach of the $ D^0 $'s line-of-flight
with respect to the primary vertex point (denoted DIP)
is required to be less than 60 $ \mu $m.
Because vertex separation, transverse momentum imbalance, 
and DIP correlate
strongly, we use a very loose DIP cut at this stage of the
analysis and include only DIP, of these variables, in FOM.

The sum of the squares of the momenta of the individual tracks
with respect to the $ D^0 $ momentum vector discriminates
between signal and background when normalized to the maximum
value possible for a candidate's reconstructed mass.
(This normalization is required to avoid kinematic biases
that could artificially create a signal by preferentially
increasing the acceptance in the signal region relative to
acceptance nearby in mass.)
This ratio is required to be greater than 0.2 for the SEED4
candidates and greater than 0.3 for the SEED3 candidates.
The product of the ratios of the four daughter tracks' transverse
separations from the secondary vertex relative to their transverse
separations from the primary vertex 
is required to be less than 0.005 for SEED4 and SEED3 candidates.
The maximum ratio for a single track is required to be
less than 1.0 and is included in FOM for SEED3 candidates; this is
unnecessary for SEED4 candidates because their distribution was cleaner upon
initial selection.

To avoid problems due to
congestion near the primary vertex, we also found it
useful to require either an absolute separation of the 
primary and secondary vertex
candidates or that the secondary vertex be ``isolated" from 
other tracks by requiring that all other tracks pass 
at least a minimum
distance from the secondary vertex candidate.
The Monte Carlo simulation fails to describe the distribution
of additional tracks in the events sufficiently well, so we
base these cuts, and the associated efficiencies, on studies
of the real $ D^0 \to K^- \pi^- \pi^+ \pi^+ $ data.
For SEED4 candidates we require either vertex separation
greater than 0.5 cm or secondary vertex isolation
greater than 20 $\mu $m.
For SEED3 candidates we require either vertex separation
greater than 0.5 cm or secondary vertex isolation
greater than 80 $\mu $m.

Our particle identification algorithm compared the light
observed in the two multicell threshold \v{C}erenkov
detectors
with that expected for the five hypotheses
$ e $, $ \mu $, $ \pi $, $ K $, and $ p $ 
for each track.
It then assigned
a probability to each hypothesis,
including {\em a priori} likelihoods for each species,
so that the sum of probabilities for
each track added to unity.
Tracks that are $ \pi $/$ K $ ambiguous had $ K $ probabilities
near 0.13.
Similarly, those that are $ K $/$ p $ ambiguous had $ K $
probabilities near 0.80.
Each kaon candidate is required to have $ K $ probability
greater than 0.20. 
The three $ K $ probabilities
are also included in FOM as independent variables.
Our Monte Carlo simulation of the \v{C}erenkov identification
does not match our data's variation with transverse
momentum well;
furthermore, the Monte Carlo distributions of $ K $
probability for the kaons in $ D^0 \to K^- K^- K^+ \pi^+ $
and $ D^0 \to K^- \pi^- \pi^+ \pi^+ $ decays differ.
In calculating the \v{C}erenkov probabilities' contributions
to FOM, for each range of $ K $ probability
we use the average of the fraction  found in the
$ D^0 \to K^- K^- K^+ \pi^+ $ Monte Carlo simulation
and in the real data's  $ D^0 \to K^- \pi^- \pi^+ \pi^+ $
signal.
We also considered using the product of the three $ K $ probabilities
in FOM, but found that
the greatest sensitivity could be achieved using them independently.
Because most tracks are pions, we found no benefit in using
\v{C}erenkov identification for the pions.

Two more variables provide some discriminating power between signal
and background.
In the Monte Carlo simulation, there were no SEED4 signal events
with proper decay time greater than 3.5 ps; in the data,
background was observed in this region.
So we removed SEED4 events with proper decay time greater than
3.5 ps.
SEED3 signal events extended past 3.5 ps, but we could not find
a cut that improved the sensitivity. 
For  both SEED4 and SEED3 events, we found that the distribution
of the cosine of the polar angle of the
sphericity axis of the candidate relative
to the candidate's line-of-flight discriminated between signal and background,
primarily because background accumulates preferentially at
values of the cosine near one.
This correlates with the scaled, summed transverse momentum
squared being small.
Having made an absolute cut on the latter quantity,
the cosine of the sphericity angle is included in FOM
for both the SEED4 and SEED3 samples.

To create FOM variables, we divided the distribution for any one
variable into four or five ranges and determined the fraction of 
signal that appeared in each range. 
Similarly, we determined from the data what fraction of background
appeared in each range.
We use $ {\cal S}_{A,i} $ to denote the probability
that a signal event falls in range $ i $ for variable $ A $.
We use $ {\cal B}_{A,i} $ for background similarly.
For example, if three variables $ {A} $, $ {B} $,
and $ { C }$ are used to define FOM,
and they are  observed
in ranges $ i $, $ j $, and $ k $ respectively,
we calculate:
\begin{equation}
{\rm FOM} = {
 { {\cal S}_{A,i} {\cal S}_{B,j} {\cal S}_{C,k} }
\over
 { {\cal B}_{A,i} {\cal B}_{B,j} {\cal B}_{C,k} } 
 } \ .
\end{equation}
This is effectively a ratio of likelihoods constructed as a
product of independent relative probabilities.
Assuming the variables  $ A $, $ B $, and $ C $ are 
statistically independent, an event's FOM is the relative signal-to-background
ratio in the $ {i,j,k} $ cell of $ {A, B,C} $ phase-space:
the signal-to-background ratio in that cell will be the
signal-to-background ratio before FOM cuts times FOM.
In  selecting events using FOM, we accept events with
FOM greater than a given value and reject those with lower values.

\begin{figure}[!]
   \begin{minipage}{6.0in}
    \centering
    \includegraphics[width=5.8in]{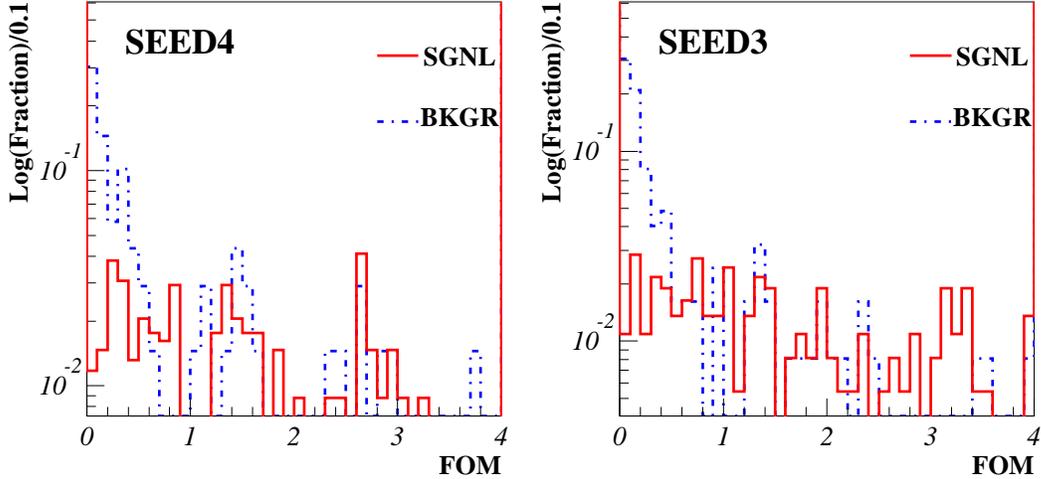}
    \caption{Figure-of-merit (FOM) distributions for Monte Carlo
     signal (solid lines) and for background (dashed lines) for
     SEED4 candidates (on the left) and SEED3 candidates (on the
     right), after all cuts on individual variables, as discussed
     in the text.
     }
    \label{fomData}
   \end{minipage}
\end{figure}\medskip

The FOM distributions for SEED4 and SEED3 Monte Carlo signal
and for real data in the sidebands (after all the non-FOM cuts)
are shown in Fig.\  \ref{fomData}.
The background accumulates preferentially at lower values of
FOM, while the signal populates the FOM distribution much
more uniformly.
In determining where to cut on FOM, we calculate the expected $ K K K \pi $
signal using (i) our observed $ K^- \pi^- \pi^+ \pi^+ $ signal,
(ii) the ratio of decay rates previously reported by E687\cite{e687},
and (iii) the relative reconstruction efficiencies determined from
our Monte Carlo simulations.
We calculate the background expected in a 15 MeV/$c^2$
window by extrapolating
the $ K^- K^- K^+ \pi^+ $ rate from
outside our masked-off range (1845 - 1885 MeV/$c^2$).
With no FOM cuts,
these assumptions predict 6.1 SEED4 and 7.5 SEED3
signal events with 4.0 and 7.2 background events, respectively.
Adding these together predicts $ S/B = 1.2 $, $ S/\sqrt{B} 
= 4.1 $, and $ S/\sqrt{S+B} = 2.7 $.
Our final selection of cuts balances our interests in maximizing
the  $ S/\sqrt{S+B} $, maximizing $ S / \sqrt{B} $,
and maintaining good $ S/B $ ratios in the SEED4 and SEED3
samples.
Our final selection of cuts is $ {\rm FOM} > 0.5 $ for SEED4
candidates and $ {\rm FOM} > 1.0 $ for SEED3 candidates.
With these cuts our algorithm predicts 11.5 signal events
and 3.1 background, giving $ S/B = 3.7 $, $ S /\sqrt{B} = 6.5 $,
and $ S/\sqrt{S+B} = 3.0 $.

Several points deserve emphasis.
The technique for selecting cuts is almost unbiased.
The data in the $ D^0 \to K^- K^- K^+ \pi^+ $
signal region have not been examined, so we avoid choices which
are subconsciously chosen to either enhance the signal level or
increase the sensitivity of an upper limit should no
signal be observed.
The potential bias in selecting cuts while looking at the
background outside the signal region is small;
this will be quantified when we discuss systematic errors.
Using FOM to combine variables that discriminate between
signal and background allows us to create a relatively robust
variable, $ S /\sqrt{S+B} $, which correlates  with our 
ability to measure the decay rate and varies slowly with changes
in FOM cuts.
We can choose looser and tighter cuts for which $ S/ B $
will vary substantially, but $ S/ \sqrt{S+B} $ should not.
This allows us to examine the data {\em a posteriori} to
identify potential problems with the analysis.

\medskip

\leftline{\bf Branching Ratio Measurement}

\begin{figure}[!]
   \begin{minipage}{2.95in}
    \centering
    \includegraphics[width=2.95in]{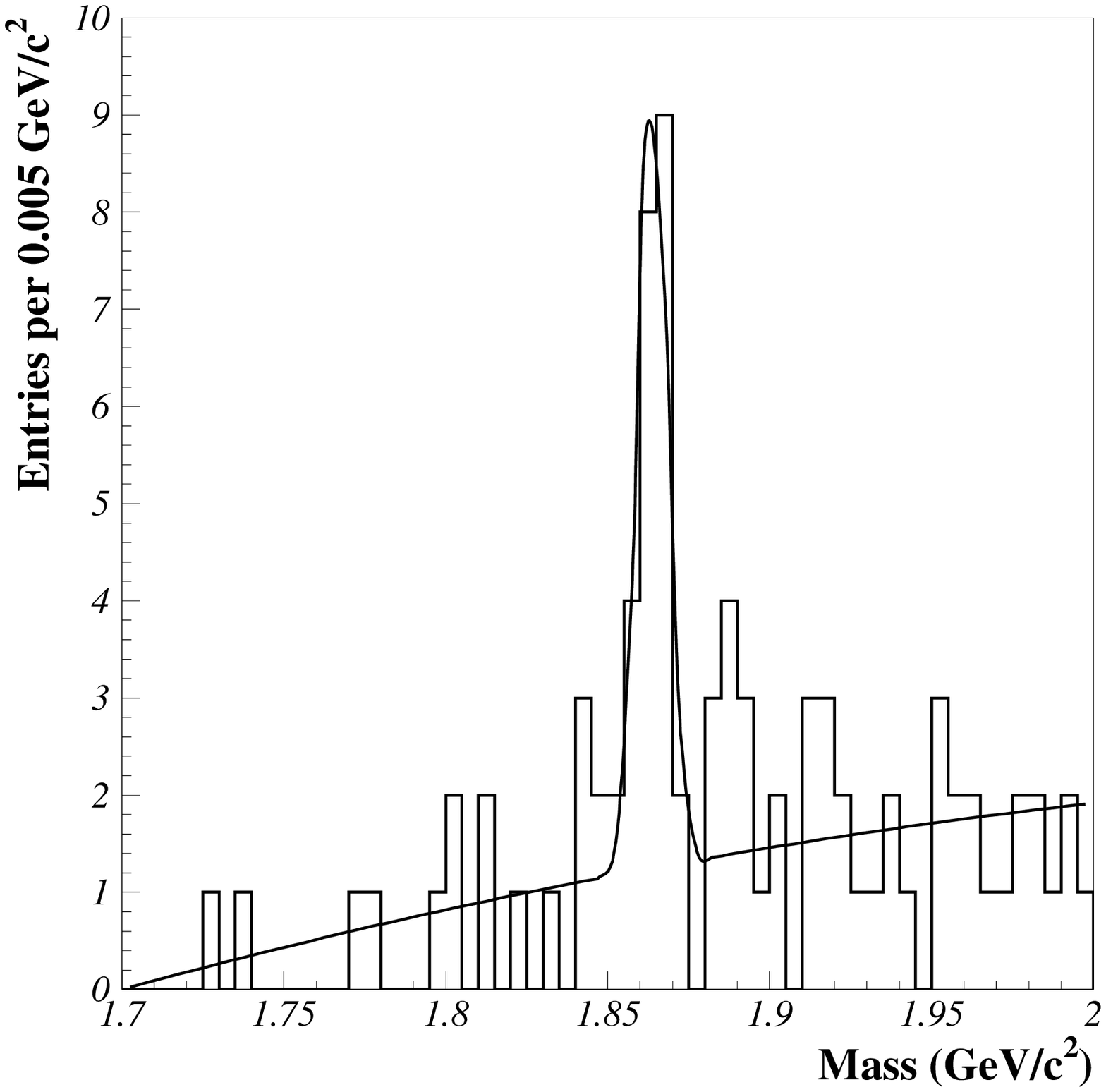}
    \caption{The $ K K K \pi $ invariant mass distribution
     for events satisfying the final selection criteria. The SEED4
     and SEED3 samples have been combined. The fitted signal level is 
     $ 18.4 \pm 5.3 $ events.}
    \label{kkkpidata}
   \end{minipage}
\hspace{0.2in}
   \begin{minipage}{2.95in}
    \centering
    \includegraphics[width=2.95in]{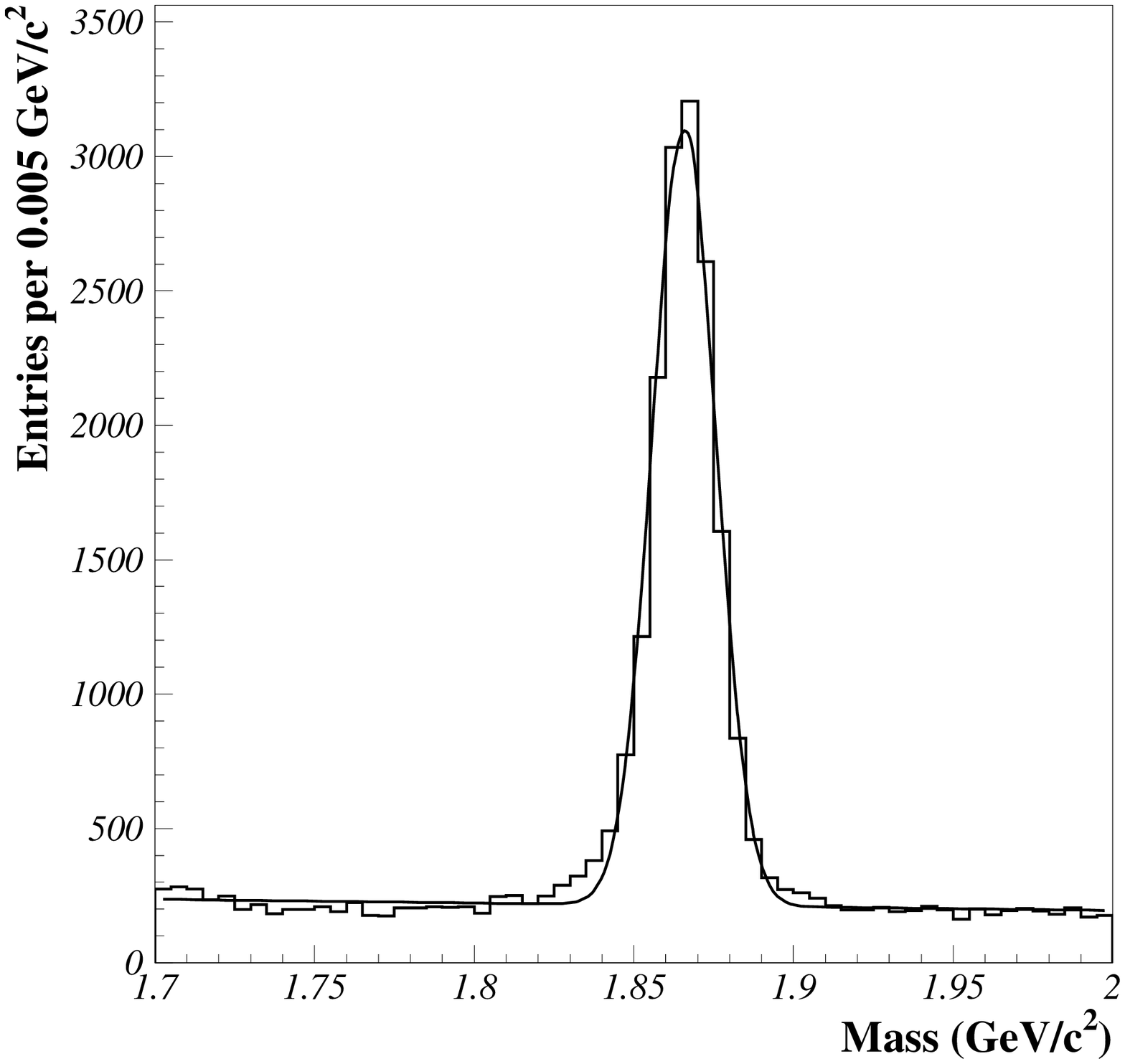}
    \caption{The $ K \pi \pi \pi $ invariant mass distribution
     for events satisfying the selection criteria described in the
     text. These criteria are similar to those used for the $ K K K \pi $
     candidates to reduce the systematic uncertainties in determining
     the relative branching ratio.
     The fitted signal level is  $ 14472 \pm 134 $ events.}
    \label{k3pidata}
   \end{minipage}
\end{figure}\medskip

{\renewcommand{\baselinestretch}{1.5}
\begin{table}[tb]
\begin{center}
\begin{tabular}{||c|c||c|c|c||}
  \cline{3-5} \cline{3-5}
  \multicolumn{2}{c||}{ } & Signal & Mass & Width \\
  \hline \hline
  \raisebox{-.8ex}{real}  & $K K K \pi$ &  18.4 $\pm$ 5.3 &
     1.8639 $\pm$ 0.0015  & 0.0045 $\pm$ 0.0014 \\
  \raisebox{+.8ex}{data}  & $K \pi \pi \pi$ & 14472 $\pm$ 134 &
     1.8658 $\pm$ 0.0001  & 0.0100 $\pm$ 0.0001 \\
  \hline
  \raisebox{-.8ex}{Monte} & $K K K \pi$ &  595 $\pm$ 26  &
      1.8646 $\pm$ 0.0002 & 0.0041 $\pm$ 0.0001 \\
  \raisebox{+.8ex}{Carlo} & $ K \pi \pi \pi$ & 2156 $\pm$ 48 &
      1.8644 $\pm$ 0.0002 & 0.0082 $\pm$ 0.0002 \\
  \hline \hline
\end{tabular}
\end{center}
\vspace{-1em}
\renewcommand{\baselinestretch}{1.2}
\caption{
Parameters determined by fitting the final real data and
Monte Carlo $ K K K \pi $ and $ K \pi \pi \pi $ samples
as described in the text.
Each Monte Carlo sample was generated with 500,000 events.
The errors quoted are statistical only.}
\label{fitResults}
\end{table}
}

The $ K^- K^- K^+ \pi^+ $ invariant mass distribution for events
satisfying the final set of cuts described above is shown in 
Fig.\  \ref{kkkpidata}, and the $ K^- \pi^- \pi^+ \pi^+ $
invariant mass distribution used for normalization is shown
in Fig.\  \ref{k3pidata}.
The cuts used for the normalization sample correspond closely to
those used for the  $ K^- K^- K^+ \pi^+ $ sample without
FOM cuts. 
The detailed requirements for finding the vertex outside
the target foils or other solid material, and an additional
requirement that the 
$ K^- \pi^- \pi^+ \pi^+ $ daughter tracks not point back to the
primary vertex, differ slightly because the $ Q $-values 
of the two decays
(summed kinetic energies
of the decay products) 
differ substantially.
Parameters for the $ K^- K^- K^+ \pi^+ $ invariant mass distribution 
are determined using an unbinned
maximum likelihood fit in which the signal is described as a Gaussian
with the mass and width allowed to float and the
background is described as a quadratic function.
Parameters for the  $ K^- K^- K^+ \pi^+ $ Monte Carlo data as well
as for the $ K^- \pi^- \pi^+ \pi^+ $ real and Monte Carlo data are
determined using binned maximum likelihood fits in which
the signals are described as Gaussian distributions with
masses and widths allowed to float and the backgrounds are
described as linear functions.
The quadratic term found in fitting the $ K^- K^- K^+ \pi^+ $ 
is small, but we allow this extra degree of freedom to
be conservative.
The Monte Carlo data has essentially no background, and adding a
quadratic term to the $ K^- \pi^- \pi^+ \pi^+ $ fit makes a negligibly
small difference, so for the Monte Carlo and normalization 
samples we present the results 
of fits with only
linear background terms.
The results of these fits are summarized in Table \ref{fitResults}.
The $ K^- K^- K^+ \pi^+ $ signal level is
$ 18.4 \pm 5.3 $ events.

To convert this signal level into a ratio of decay rates we need
the $ K^- \pi^- \pi^+ \pi^+ $ signal level ($14472 \pm 134 $ events)
and the relative efficiency for the two decay modes.
The latter differs from unity for three reasons. 
First, the $ Q $-value for the
$ K K K \pi $ decay is smaller than that for the $ K \pi \pi \pi $ 
decay. 
This leads to very different track opening angles, and
hence to very different vertex resolutions.
As a result, vertex reconstruction efficiencies and vertex separation
distributions differ. 
Second, the background in the  $ K K K \pi $ sample is reduced using FOM
cuts, a procedure not necessary in the  normalization sample.
Finally, the $ K K K \pi $ sample has two additional kaons,
which reduces the particle identification efficiency.
We start with the relative efficiency determined from the Monte Carlo
simulations, $ 0.275 \pm 0.013 $, (where the reported error is
the statistical uncertainty from the Monte Carlo samples) and make corrections
to account for differences between real data and Monte Carlo data observed using the
$ D^0 \to K^- \pi^- \pi^+ \pi^+ $ signal.
These corrections are summarized in Table \ref{corrections}.
Taken together, we estimate the efficiency for
$ D^0 \to K^- K^- K^+ \pi^+ $ relative to that for $ D^0 \to
K^- \pi^- \pi^+ \pi^+ $ to be 15\% less than that determined directly
from the Monte Carlo simulations.

{\renewcommand{\baselinestretch}{1.2}
\begin{table}[htb]
\vspace{2em}
\begin{center}
\begin{tabular}{||c|c|c||}
  \hline \hline
  systematic variation due to         & raise $\epsilon_{KKK \pi_{rel}}$ by
                                      & correction factor \\
  \hline \hline
  kaon \v{C}erenkov efficiency           & $-18\%$  & 0.82  \\
  SEED4 and SEED3 fractions           & $+2.5\%$ & 1.025 \\
  vertex separation                       & $+1.1\%$ & 1.011 \\
  $p_T$-balance                            & $+0.5\%$ & 1.005 \\
  \hline \hline
  total correction factor             & $-15\%$  & 0.85  \\
  \hline \hline
\end{tabular}
\end{center}
\caption[Summary of corrections to the relative efficiency from Monte
         Carlo for reconstructing $ K K K \pi $ and 
         $ K K K \pi $ final states.]
        {Summary of corrections to the relative efficiency from Monte
         Carlo for reconstructing $ K K K \pi $ and 
         $ K \pi \pi \pi $ final states.
         The relative efficiency used for determining the relative
         branching ratio, $\Gamma_{K K K \pi } /\Gamma_{K \pi \pi \pi}$, will be
         0.85 times that found from the Monte Carlo. The total correction
         factor has been calculated by multiplying the individual
         correction factors.}
\label{corrections}
\end{table}
}

The significant sources of systematic uncertainty in the
ratio of decay rates are
summarized in Table \ref{systematics}.
Each of the correction factors described above has a
corresponding uncertainty determined by
studying the $ D^0 \to K^- \pi^- \pi^+ \pi^+ $ data
that pass cuts, and the Monte Carlo samples for both decay modes.
The Monte Carlo statistics contribute a $ 4.7 \% $
uncertainty.
Systematic differences in tracking and vertexing efficiencies between
real data and Monte Carlo samples have been 
studied previously\cite{e791_previous}.
They contribute an additional 5\% systematic uncertainty beyond
that determined for the correction
factors; this is effectively an uncertainty on the relative
efficiency for the loosest cuts used.
We vary the fit used to extract the number of
$ D^0 \to K^- K^- K^+ \pi^+ $ signal events allowing
both linear and quadratic backgrounds, fixed and floating masses,
and fixed and floating Gaussian width, and all possible combinations.
The number of signal events ranges from $ 17.6 \pm 4.7 $ to $ 18.8 \pm 5.2 $.
Each fit describes the data adequately --
for each fit the $ \chi^2 $ per degree of freedom is less than one.
We investigated backgrounds due to misidentified charm decays using
Monte Carlo simulations
and found the overall shape to agree well with that found in the fits.
We could associate a systematic uncertainty of $ 4.3 \% $
with our fitting procedure; this would cover the largest excursion of the
fit results from the central value reported.
Because the background near the signal region may be higher
than predicted by our fit over the whole range shown,
we considered several piece-wise linear fits
as well. 
We estimate an additional one event systematic uncertainty
in the background level which we add in quadrature to give
a total systematic uncertainty of
$ 7.6 \% $ due to the shapes of the signal and background.

{\renewcommand{\baselinestretch}{1.0}
\begin{table}[htb]
\vspace{1em}
\begin{center}
\begin{tabular}{||lr|ll||}
  \hline \hline
  \multicolumn{2}{|| l|}{differences between Monte Carlo}      &       & \\
  \multicolumn{2}{|| l|}{and real data (after corrections):} 
                &       & \\
                & kaon \v{C}erenkov efficiency           &  10\% & \\
                & tracking and vertexing efficiencies &   5\% & \\
                & SEED4 and SEED3 fractions           & 1.8\% & \\
                & vertex seapartion requirement       & 1.1\% & \\
                & $ p_T $-balance                     & 0.5\% & \\
                & sub total                           & & 11.4\% \\ \hline
  \multicolumn{2}{|| l |}{Statistical fluctuations in Monte} &       & \\
  \multicolumn{2}{|| l |}{Carlo data, including difference} &  &  \\ 
   \multicolumn{2}{|| l |}{between non-resonant and $ \phi K \pi $} 
        &  & 4.7\% \\ \hline
  \multicolumn{2}{|| l |}{Signal and background shapes}    & & 7.6\% \\ \hline
  \multicolumn{2}{|| l |}{FOM predictions}                & & 1.5\% \\ \hline
  \hline 
  \multicolumn{2}{|| l |}{Total (added in quadrature)}    & & 14.6\% \\
  \hline \hline
\end{tabular}
\end{center}
\renewcommand{\baselinestretch}{1.0}
\caption{Summary of systematic errors. The total has been calculated by
         adding the individual contributions in quadrature.}
\label{systematics}
\end{table}
}

With the corrected relative efficiency described above, and adding
the systematic uncertainties in quadrature, the ratio of
decay rates is:
\begin{equation}
      \frac{\Gamma(D^0\to K^-K^- K^+\pi^+)} 
{\Gamma (D^0\to K^-\pi^-\pi^+\pi^+)}  =   (0.54 \pm 0.16 \pm 0.08)
\times 10^{-2} \ .
\label{relBR}
\end{equation}
The first error is statistical; the second is systematic.
Using the $ D^0 \to K^- \pi^- \pi^+ \pi^+ $ branching ratio reported by the
Particle Data Group\cite{pdg}, $ (7.6 \pm 0.4) \% $,
and folding its error into our final systematic uncertainty, we obtain
\begin{equation}
{\cal BR} ( D^0 \to K^- K^- K^+ \pi^+ ) = (4.1 \pm 1.2 \pm 0.6 ) \times 10^{-4} \ .
\end{equation}
The ratio 
of decay rates reported by E687 is $ ( 0.28 \pm 0.07 \pm 0.01) \% $\cite{e687}.
The difference between our result and the E687 result is $ ( 0.26 \pm 0.19) \% $.
The fractional errors for the two results are 0.32 (this work) and 0.25 (E687)
where the statistical and systematic errors have been added in quadrature.

\bigbreak
\newpage
\leftline{\bf Search for Resonant Substructure}

\begin{figure}[!]
   \begin{minipage}{2.95in}
    \centering
    \includegraphics[width=2.95in]{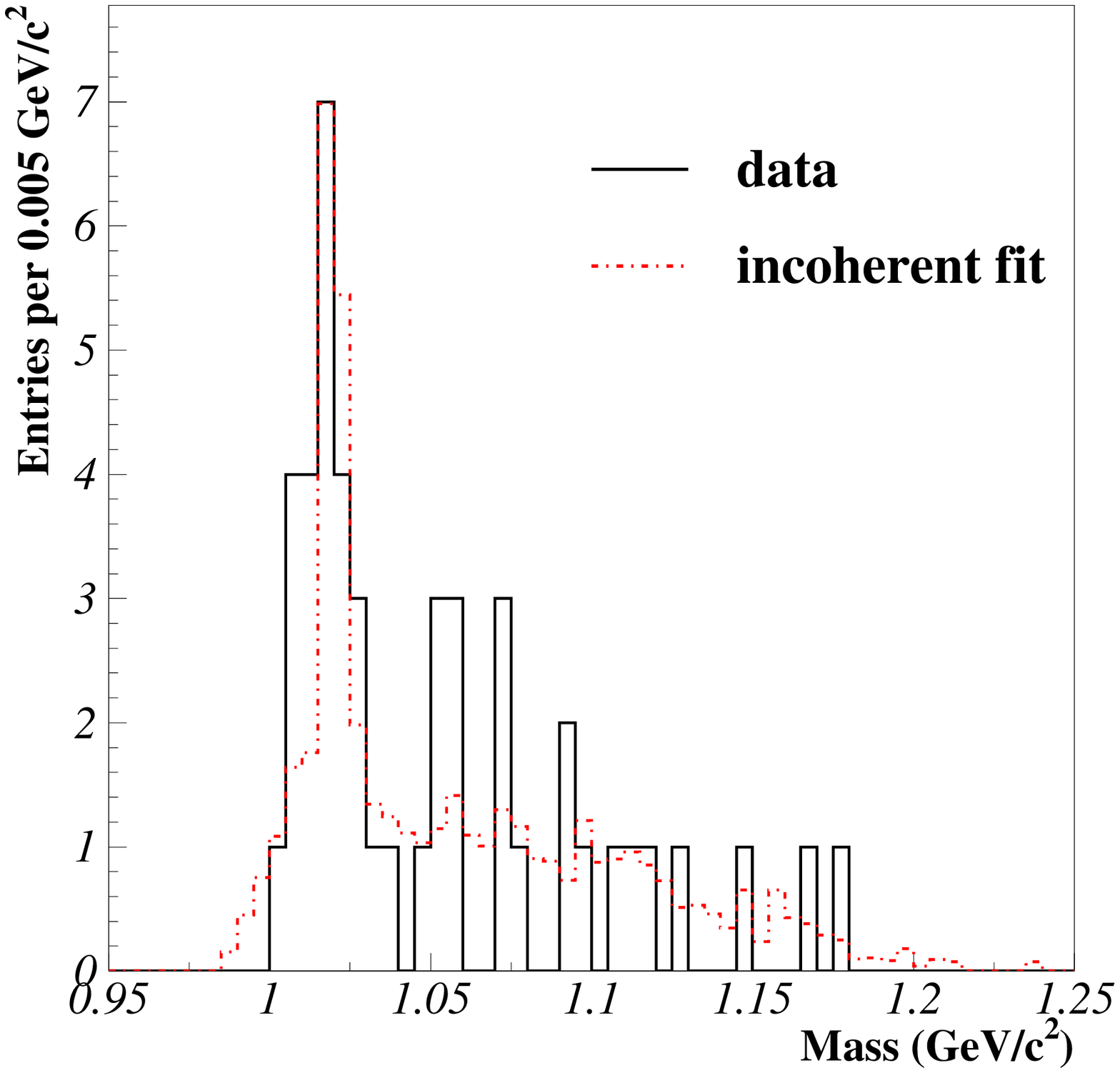}
    \caption{
     $ K^- K^+  $ invariant mass distributions for
     candidates with  1.855 GeV/$c^2$ $ < $   $ m( K^- K^- K^+ \pi^+ ) $
     $ < $ 
     1.875 GeV/$c^2$. 
     There are two entries per  $ D^0 $ candidate.
     The solid line histogram is the real data.
     The  dashed line histogram is a toy model in which the signal fraction
     is described as 70\% from $ \phi K \pi $ and 30\% non-resonant,
     and the background scaled from the data,
     as discussed in the text.}
    \label{KKSignal}
   \end{minipage}
   \qquad
   \begin{minipage}{2.95in}
    \centering
    \includegraphics[width=2.95in]{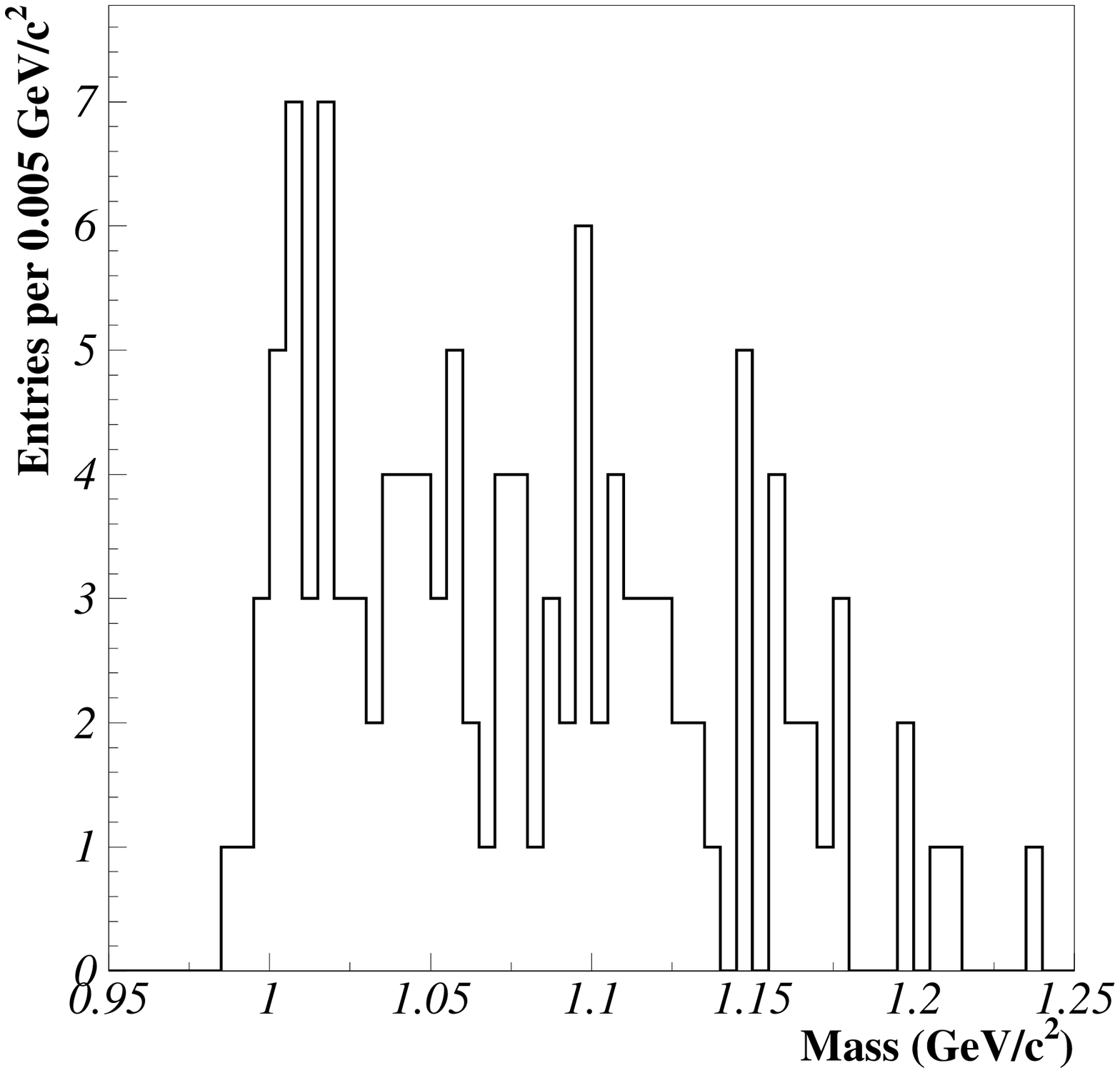}
    \caption{$ K^- K^+  $ invariant mass distribution for
     candidates  in the background region.
     There are two entries per $ K^- K^- K^+ \pi^+ $ candidate.}
    \label{KKBkgd}
   \end{minipage}
\end{figure}

Two or more of the final state hadrons in a $ D^0 \to K^- K^- K^+ \pi^+ $
final state might be the decay products of an intermediate resonance.
The signal observed in this experiment is small, and the phase-space so small
that it will distort the shapes of broad resonances that appear as intermediate
states.
Hence, we have not attempted a coherent amplitude analysis similar to 
the analysis we
did for the decay $ D^0 \to K^- K^+ \pi^- \pi^+ $\cite{lalith} or 
similar to the incoherent amplitude analysis done by MARK III for the
decay $ D^0 \to K^- \pi^- \pi^+ \pi^+ $\cite{markIII}.
Rather, we have looked only at $ K^- K^+ $ invariant mass distributions
of $D^0 \to  K^- K^- K^+ \pi^+ $
candidates (two pairs per candidate).
The signal distribution 
(for events with  1.855 GeV/$c^2$ $ < $   $ m( K^- K^- K^+ \pi^+ ) $ 
$ < $ 
1.875 GeV/$c^2$),
seen in Fig.\  \ref{KKSignal}, shows an accumulation of entries
near the $ \phi $ mass.
In comparison, the corresponding distribution
for background events in the ranges $ 1.700 $ GeV/$c^2 < m( K K K \pi ) < 1.845 $ GeV/$c^2 $
and $ 1.885 $ GeV/$c^2 < m( K K K \pi) < 2.000 $ GeV/$c^2$,
seen in Fig.\  \ref{KKBkgd}, has
a much broader distribution, with only a hint of any structure 
at low mass.

\begin{figure}[t]
   \begin{minipage}{2.8in}
    \centering
    \includegraphics[width=2.8in]{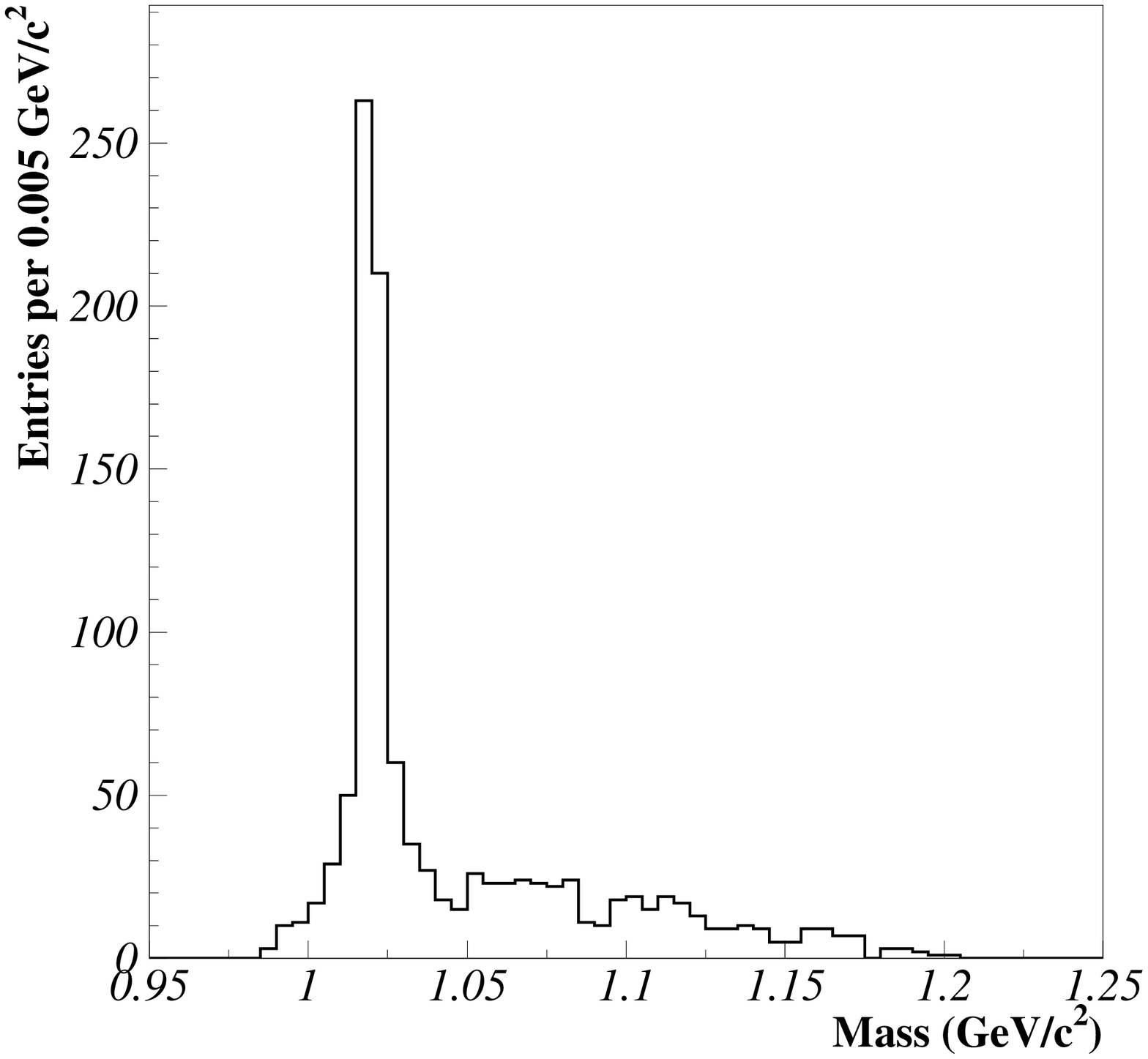}
    \caption{$ K^- K^+ $ invariant mass distribution for
     $ D^0 \to \phi K \pi $ Monte Carlo events. There are two entries
     per $ D^0 $ candidate.}
    \label{PhiMC}
   \end{minipage}
\hspace{0.2in}
   \begin{minipage}{2.8in}
    \centering
    \includegraphics[width=2.8in]{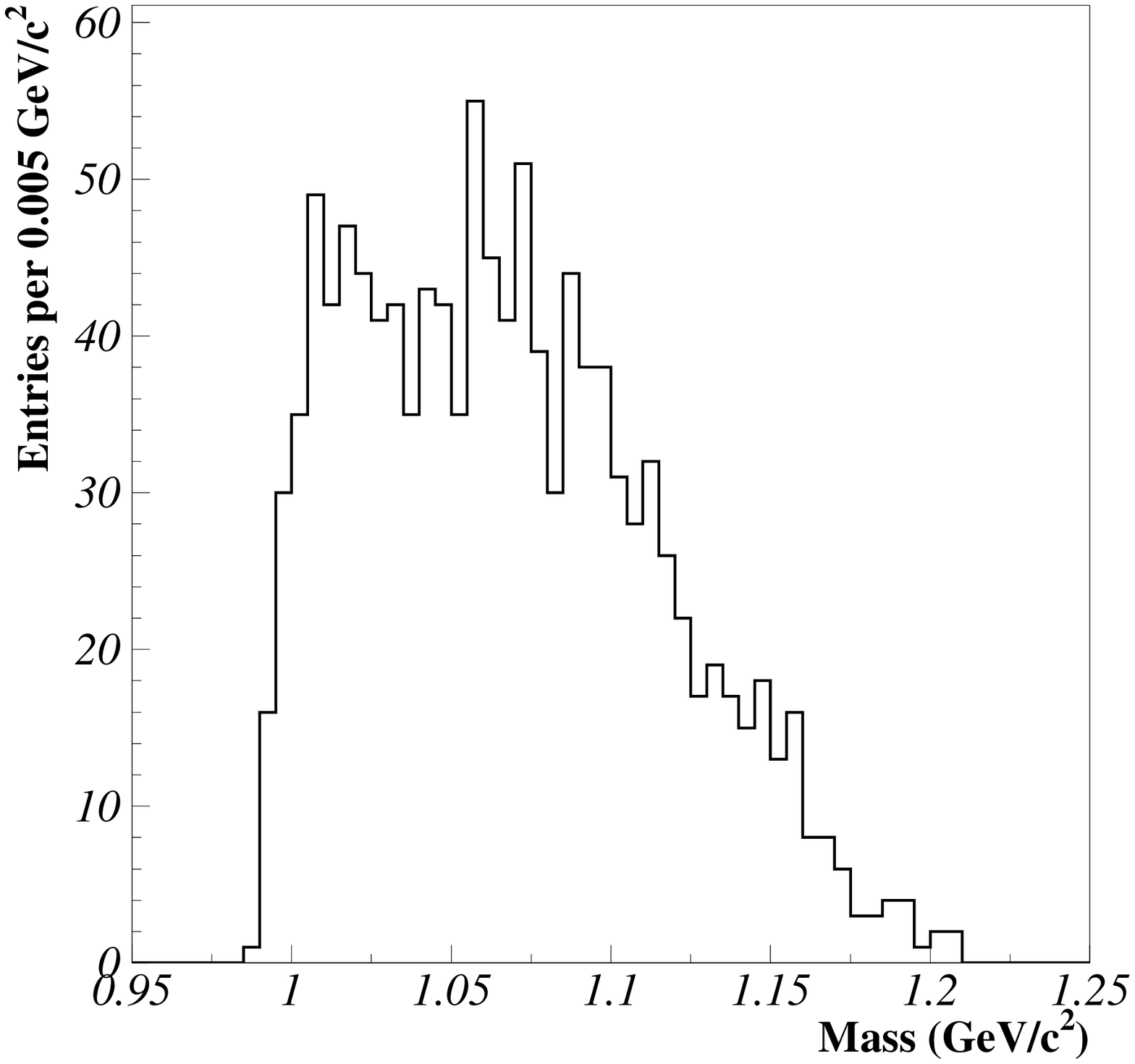}
    \caption{$ K^- K^+  $ invariant mass distribution for
     non-resonant 
    $ D^0 \to K^- K^- K^+ \pi^+ $
    Monte Carlo events. There are two entries
    per  $ D^0 $ candidate.}
    \label{NRMC}
   \end{minipage}
\end{figure}\medskip

To understand the nature of our signal better, we generated two Monte
Carlo samples. 
In our non-resonant $ D^0 \to K^- K^- K^+ \pi^+ $
simulation, the generated events populate four-body phase-space uniformly.
In our $ D^0 \to \phi K^- \pi^+ $; $ \phi \to K^- K^+ $ simulation,
the generated events populate the ($ \phi ,  K ,  \pi $)
three-body phase-space
uniformly.
Both Monte Carlo samples are fully simulated and then reconstructed 
and analyzed as
were the real data.
The $ K^- K^+ $ invariant mass distributions corresponding to those for
the real data are shown in Figs.\ \ref{PhiMC} and \ref{NRMC}.
The two distributions differ qualitatively.

Without trying to do a real amplitude analysis,
we fit the $ K^- K^+ $ invariant mass distribution 
of Fig.\  \ref{KKSignal} as an incoherent
sum of the shapes of the two Monte Carlo models and of
the background region.
We use a binned maximum likelihood fit in which
the two signal fractions float freely and the background
fraction floats, but we add a term to the likelihood function
to account for the difference between the background fraction and that
determined from the earlier fit of the $ K K K \pi $ used
for the branching ratio measurement.
This fit
(which is superposed on the data in Fig.\  \ref{KKSignal})
finds that $ 0.7 \pm 0.3 $ of the  $ K K K \pi $ signal
comes from $ \phi K \pi $ decay.
This result
indicates that intermediate $ \phi $ production is an
important mechanism in $ D^0 \to K^- K^- K^+ \pi^+ $ decay.

\leftline{\bf Interpretation of the Relative Branching Ratio}

The relative branching ratio determined in Eq. \ref{relBR} is small
primarily because the $ Q $-value of the $ K K K \pi $ decay is
much less than that of the $ K \pi \pi \pi $ decay.
The phase space
for non-resonant four-body $ K K K \pi $ decay, $ \Omega_{K K K \pi} $, is only
$ 1.43 \times 10^{-2} $ times that for non-resonant four-body $ K \pi \pi \pi $ decay, 
$ \Omega_{K \pi \pi \pi} $.
If one assumes that both decays are purely non-resonant,
the ratio of invariant matrix elements,
$ {\cal R} $, is a constant and can be determined using
\begin{equation}
{ {\Gamma_{K K K \pi}} \over { \Gamma_{K \pi \pi \pi } } } =
{ {\Omega_{K K K \pi}} \over { \Omega_{K \pi \pi \pi } } } \times {\cal R}
\end{equation}
to be $ {\cal R} = 0.32 \pm 0.10 $.
\thicklines

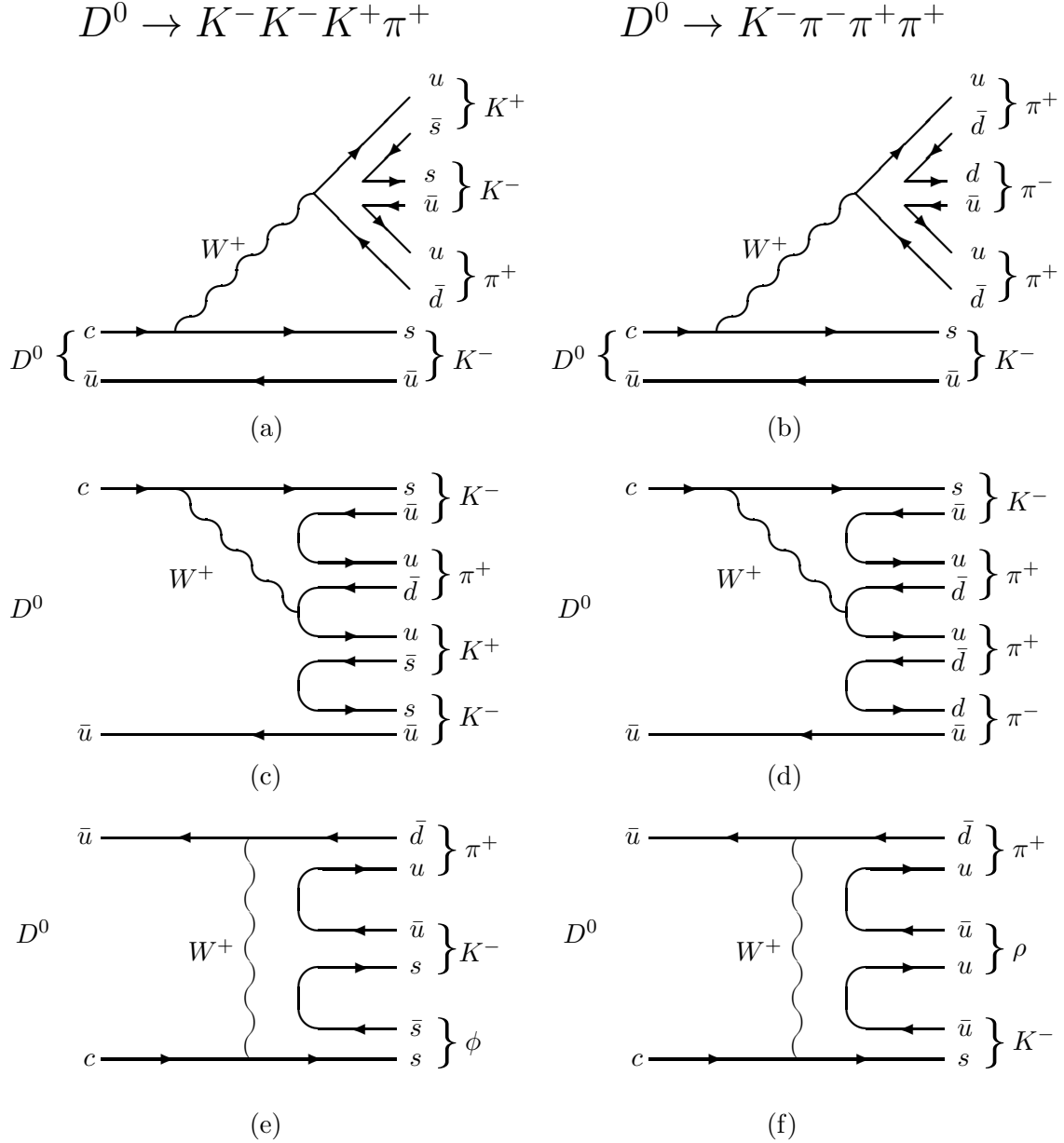
\begin{figure}[htbp]
\begin{center}

\vspace{2em}
\begin{picture}(43000,12000)

  \Xone=4000
  \Yone=-1000
  \Xtwo=12000  
  
  {\Large
     \put( 3000,13000){$D^0 \rightarrow K^-K^-K^+\pi^+$}
     \put(25000,13000){$D^0 \rightarrow K^-\pi^-\pi^+\pi^+$}
  }

  \drawline\fermion[\E\REG](\Xone,\Yone)[\Xtwo]

  \Xthree=\pfrontx
  \Xfour=\Xtwo
  \global\divide\Xfour by 2
  \global\advance\Xthree by \Xfour
  \drawarrow[\W\ATTIP](\Xthree,\pfronty)
  
  \Xthree=\pfrontx
  \global\advance\Xthree  by -750
  \global\advance\pfronty by -250
  \put(\Xthree,\pfronty){$\bar{u}$}
  {\LARGE 
    \global\advance\pfrontx by -2000
    \global\advance\pfronty by   750
    \put(\pfrontx,\pfronty){\{}
  }
  \global\advance\pfrontx by -1750
  \put(\pfrontx,\pfronty){$D^0$}

  \Xfour=\pbackx
  \global\advance\Xfour  by  250
  \global\advance\pbacky by -250
  \put(\Xfour,\pbacky){$\bar{u}$}
  
  \global\advance\Yone by 2000
  \drawline\fermion[\E\REG](\Xone,\Yone)[\Xtwo]
    
  \Xthree=\pfrontx
  \Xfour=\Xtwo
  \global\divide\Xfour by 6
  \global\advance\Xthree by \Xfour  
  \drawarrow[\E\ATTIP](\Xthree,\pfronty)
  
  \Xthree=\pfrontx
  \Xfour=\Xtwo
  \global\divide\Xfour by 3
  \global\multiply\Xfour by 2
  \global\advance\Xthree by \Xfour  
  \drawarrow[\E\ATTIP](\Xthree,\pfronty)

  \Xthree=\pfrontx
  \Xfour=\pbackx
  \global\advance\Xthree  by -750
  \global\advance\pfronty by -250
  \put(\Xthree,\pfronty){$c$}

  \global\advance\Xfour  by  250
  \global\advance\pbacky by -250
  \put(\Xfour,\pbacky){$s$}
  {\LARGE 
    \global\advance\pbackx by  1000
    \global\advance\pbacky by -1250
    \put(\pbackx,\pbacky){\}}
  }
  \global\advance\pbackx by  1250
  \put(\pbackx,\pbacky){$K^-$}


  \Xthree=\pfrontx
  \Xfour=\Xtwo
  \global\divide\Xfour by 4
  \global\advance\Xthree by \Xfour  
  \drawline\photon[\NE\REG](\Xthree,\pmidy)[9]
  \global\advance\Xthree by 1000
  \Ythree=\Yone
  \global\advance\Ythree by 3000
  \put(\Xthree,\Ythree){$W^+$}

  \Xthree=\Xtwo
  \global\divide\Xthree by 3
  \global\advance\Xthree by 1500
  \drawline\fermion[\NE\REG](\photonbackx,\photonbacky)[\Xthree]
  \drawarrow[\NE\ATTIP](\pmidx,\pmidy)
  \global\advance\pbackx by 750 
  \global\advance\pbacky by 500 
  \put(\pbackx,\pbacky){$u$}

  {\LARGE 
    \global\advance\pbackx by  1000
    \global\advance\pbacky by -1250
    \put(\pbackx,\pbacky){\}}
  }
  \global\advance\pbackx by  1250
  \put(\pbackx,\pbacky){$K^+$}

  \drawline\fermion[\SE\REG](\photonbackx,\photonbacky)[\Xthree]
  \drawarrow[\NW\ATBASE](\pmidx,\pmidy)
  \global\advance\pbackx by  750
  \global\advance\pbacky by -750
  \put(\pbackx,\pbacky){$\bar{d}$}

  {\LARGE 
    \global\advance\pbackx by  1000
    \global\advance\pbacky by   750
    \put(\pbackx,\pbacky){\}}
  }
  \global\advance\pbackx by  1250
  \global\advance\pbacky by   100
  \put(\pbackx,\pbacky){$\pi^+$}


  \Xsix=\photonbackx
  \Ysix=\photonbacky
  \global\advance\Xsix  by  2000
  \global\advance\Ysix  by  500

  \Xfour=\Xthree
  \Xfive=\Xthree
  \global\advance\Xfour by -2800
  \global\advance\Xfive by -3800

  \drawline\fermion[\NE\REG](\Xsix,\Ysix)[\Xfour]
  \drawarrow[\SW\ATTIP](\pmidx,\pmidy)
  \global\advance\pbackx by  750 
  \put(\pbackx,\pbacky){$\bar{s}$}

  \drawline\fermion[\E\REG](\Xsix,\Ysix)[\Xfive]
  \global\advance\pmidx by 800
  \drawarrow[\E\ATTIP](\pmidx,\pmidy)
  \global\advance\pbackx by  750 
  \put(\pbackx,\pbacky){$s$}

  {\LARGE 
    \global\advance\pbackx by  1000
    \global\advance\pbacky by  -750
    \put(\pbackx,\pbacky){\}}
  }
  \global\advance\pbackx by  1250
  \put(\pbackx,\pbacky){$K^-$}

  \Xseven=\photonbackx
  \Yseven=\photonbacky
  \global\advance\Xseven  by  2000
  \global\advance\Yseven  by -500
  \drawline\fermion[\SE\REG](\Xseven,\Yseven)[\Xfour]
  \drawarrow[\SE\ATTIP](\pmidx,\pmidy)
  \global\advance\pbackx by  750 
  \global\advance\pbacky by -250 
  \put(\pbackx,\pbacky){$u$}

  \drawline\fermion[\E\REG](\Xseven,\Yseven)[\Xfive]
  \drawarrow[\W\ATTIP](\pmidx,\pmidy)
  \global\advance\pbackx by  750 
  \global\advance\pbacky by -250 
  \put(\pbackx,\pbacky){$\bar{u}$}

  \put(10000,-3250){(a)}


  \Xone=26000
  \Yone=-1000
  \Xtwo=12000  

  \drawline\fermion[\E\REG](\Xone,\Yone)[\Xtwo]

  \Xthree=\pfrontx
  \Xfour=\Xtwo
  \global\divide\Xfour by 2
  \global\advance\Xthree by \Xfour
  \drawarrow[\W\ATTIP](\Xthree,\pfronty)
  
  \Xthree=\pfrontx
  \global\advance\Xthree  by -750
  \global\advance\pfronty by -250
  \put(\Xthree,\pfronty){$\bar{u}$}
  {\LARGE 
    \global\advance\pfrontx by -2000
    \global\advance\pfronty by   750
    \put(\pfrontx,\pfronty){\{}
  }
  \global\advance\pfrontx by -1750
  \put(\pfrontx,\pfronty){$D^0$}

  \Xfour=\pbackx
  \global\advance\Xfour  by  250
  \global\advance\pbacky by -250
  \put(\Xfour,\pbacky){$\bar{u}$}
  
  \global\advance\Yone by 2000
  \drawline\fermion[\E\REG](\Xone,\Yone)[\Xtwo]
    
  \Xthree=\pfrontx
  \Xfour=\Xtwo
  \global\divide\Xfour by 6
  \global\advance\Xthree by \Xfour  
  \drawarrow[\E\ATTIP](\Xthree,\pfronty)
  
  \Xthree=\pfrontx
  \Xfour=\Xtwo
  \global\divide\Xfour by 3
  \global\multiply\Xfour by 2
  \global\advance\Xthree by \Xfour  
  \drawarrow[\E\ATTIP](\Xthree,\pfronty)

  \Xthree=\pfrontx
  \Xfour=\pbackx
  \global\advance\Xthree  by -750
  \global\advance\pfronty by -250
  \put(\Xthree,\pfronty){$c$}

  \global\advance\Xfour  by  250
  \global\advance\pbacky by -250
  \put(\Xfour,\pbacky){$s$}
  {\LARGE 
    \global\advance\pbackx by  1000
    \global\advance\pbacky by -1250
    \put(\pbackx,\pbacky){\}}
  }
  \global\advance\pbackx by  1250
  \put(\pbackx,\pbacky){$K^-$}


  \Xthree=\pfrontx
  \Xfour=\Xtwo
  \global\divide\Xfour by 4
  \global\advance\Xthree by \Xfour  
  \drawline\photon[\NE\REG](\Xthree,\pmidy)[9]
  \global\advance\Xthree by 1000
  \Ythree=\Yone
  \global\advance\Ythree by 3000
  \put(\Xthree,\Ythree){$W^+$}

  \Xthree=\Xtwo
  \global\divide\Xthree by 3
  \global\advance\Xthree by 1500
  \drawline\fermion[\NE\REG](\photonbackx,\photonbacky)[\Xthree]
  \drawarrow[\NE\ATTIP](\pmidx,\pmidy)
  \global\advance\pbackx by 750 
  \global\advance\pbacky by 500 
  \put(\pbackx,\pbacky){$u$}

  {\LARGE 
    \global\advance\pbackx by  1000
    \global\advance\pbacky by -1250
    \put(\pbackx,\pbacky){\}}
  }
  \global\advance\pbackx by  1250
  \global\advance\pbacky by  100
  \put(\pbackx,\pbacky){$\pi^+$}

  \drawline\fermion[\SE\REG](\photonbackx,\photonbacky)[\Xthree]
  \drawarrow[\NW\ATBASE](\pmidx,\pmidy)
  \global\advance\pbackx by  750
  \global\advance\pbacky by -750
  \put(\pbackx,\pbacky){$\bar{d}$}

  {\LARGE 
    \global\advance\pbackx by  1000
    \global\advance\pbacky by   750
    \put(\pbackx,\pbacky){\}}
  }
  \global\advance\pbackx by  1250
  \global\advance\pbacky by  100
  \put(\pbackx,\pbacky){$\pi^+$}


  \Xsix=\photonbackx
  \Ysix=\photonbacky
  \global\advance\Xsix  by  2000
  \global\advance\Ysix  by  500

  \Xfour=\Xthree
  \Xfive=\Xthree
  \global\advance\Xfour by -2800
  \global\advance\Xfive by -3800

  \drawline\fermion[\NE\REG](\Xsix,\Ysix)[\Xfour]
  \drawarrow[\SW\ATTIP](\pmidx,\pmidy)
  \global\advance\pbackx by  750 
  \put(\pbackx,\pbacky){$\bar{d}$}

  \drawline\fermion[\E\REG](\Xsix,\Ysix)[\Xfive]
  \global\advance\pmidx by 800
  \drawarrow[\E\ATTIP](\pmidx,\pmidy)
  \global\advance\pbackx by  750 
  \put(\pbackx,\pbacky){$d$}

  {\LARGE 
    \global\advance\pbackx by  1000
    \global\advance\pbacky by  -750
    \put(\pbackx,\pbacky){\}}
  }
  \global\advance\pbackx by  1250
  \global\advance\pbacky by  100
  \put(\pbackx,\pbacky){$\pi^-$}

  \Xseven=\photonbackx
  \Yseven=\photonbacky
  \global\advance\Xseven  by  2000
  \global\advance\Yseven  by -500
  \drawline\fermion[\SE\REG](\Xseven,\Yseven)[\Xfour]
  \drawarrow[\SE\ATTIP](\pmidx,\pmidy)
  \global\advance\pbackx by  750 
  \global\advance\pbacky by -250 
  \put(\pbackx,\pbacky){$u$}

  \drawline\fermion[\E\REG](\Xseven,\Yseven)[\Xfive]
  \drawarrow[\W\ATTIP](\pmidx,\pmidy)
  \global\advance\pbackx by  750 
  \global\advance\pbacky by -250 
  \put(\pbackx,\pbacky){$\bar{u}$}

  \put(31000,-3250){(b)}

\end{picture}

\vspace{3em}
\begin{picture}(43000,12000)

  \Xone=4000
  \Yone=0
  \Xtwo=12000  
   
  \drawline\fermion[\E\REG](\Xone,\Yone)[\Xtwo]
 
  \Xthree=\pfrontx
  \Xfour=\Xtwo
  \global\divide\Xfour by 2
  \global\advance\Xthree by \Xfour
  \drawarrow[\W\ATTIP](\Xthree,\pfronty)
   
  \Xthree=\pfrontx
  \Xfour=\pbackx
  \global\advance\Xthree by -1000
  \global\advance\Xfour by 1000

  \global\advance\pfrontx by -1000
  \global\advance\pfronty by -250
  \put(\pfrontx,\pfronty){$\bar{u}$}
  {\LARGE
     \global\advance\pfrontx by -1250
     \global\advance\pfronty by  5000
  }
  \global\advance\pfrontx by -1500
  \put(\pfrontx,\pfronty){$D^0$}

  \global\advance\pbackx by  250
  \global\advance\pbacky by -250
  \put(\pbackx,\pbacky){$\bar{u}$}
  {\LARGE
     \global\advance\pbackx by 1000
     \global\advance\pbacky by 350
     \put(\pbackx,\pbacky){\}}
  }
  \global\advance\pbackx by 1250
  \global\advance\pbacky by  250
  \put(\pbackx,\pbacky){$K^-$}

  \global\advance\Yone by 10000
  \drawline\fermion[\E\REG](\Xone,\Yone)[\Xtwo]
     
  \Xthree=\pfrontx
  \Xfour=\Xtwo
  \global\divide\Xfour by 6
  \global\advance\Xthree by \Xfour  
  \drawarrow[\E\ATTIP](\Xthree,\pfronty)
   
  \Xthree=\pfrontx
  \Xfour=\Xtwo
  \global\divide\Xfour by 3
  \global\multiply\Xfour by 2
  \global\advance\Xthree by \Xfour  
  \drawarrow[\E\ATTIP](\Xthree,\pfronty)

  \Xthree=\pfrontx
  \Xfour=\pbackx
  \global\advance\Xthree by -1000
  \global\advance\Xfour by 1000

  \global\advance\pfrontx by -1000
  \global\advance\pfronty by -250
  \put(\pfrontx,\pfronty){$c$}

  \global\advance\pbackx by  250 
  \global\advance\pbacky by -250
  \put(\pbackx,\pbacky){$s$}
  

  \Xthree=\pfrontx
  \Xfour=\Xtwo
  \global\divide\Xfour by 3
  \global\advance\Xthree by \Xfour  
  \drawline\photon[\SE\REG](\Xthree,\pmidy)[8]
  \global\advance\Xthree by -300
  \Ythree=\Yone
  \global\advance\Ythree by -4000
  \put(\Xthree,\Ythree){$W^+$}
   


  \global\advance\photonbackx by 1000
  \put(\photonbackx,\photonbacky){\oval(2000,2000)[l]}

  \Xseven=\Xtwo
  \global\divide\Xseven by 4

  \Xfour=\photonbackx
  \Yfour=\photonbacky
  \global\advance\Yfour by 1000
  \drawline\fermion[\E\REG](\Xfour,\Yfour)[\Xseven]
  \global\advance\pmidx by -800
  \drawarrow[\W\ATTIP](\pmidx,\pmidy)
  \global\advance\pbackx by  250
  \global\advance\pbacky by -500
  \put(\pbackx,\pbacky){$\bar{d}$}
  {\LARGE
     \global\advance\pbackx by 1000
     \global\advance\pbacky by 500
     \put(\pbackx,\pbacky){\}}
  }
  \global\advance\pbackx by 1250
  \global\advance\pbacky by  150
  \put(\pbackx,\pbacky){$\pi^+$}

  \Yfour=\photonbacky
  \global\advance\Yfour by -1000
  \drawline\fermion[\E\REG](\Xfour,\Yfour)[\Xseven]
  \drawarrow[\E\ATTIP](\pmidx,\pmidy)
  \global\advance\pbackx by  250
  \global\advance\pbacky by -150
  \put(\pbackx,\pbacky){$u$}


  \Ysix=\photonbacky
  \global\advance\Ysix by 3000
  \put(\photonbackx,\Ysix){\oval(2000,2000)[l]}

  \Xfour=\photonbackx
  \Yfour=\Ysix
  \global\advance\Yfour by 1000
  \drawline\fermion[\E\REG](\Xfour,\Yfour)[\Xseven]
  \global\advance\pmidx by -800
  \drawarrow[\W\ATTIP](\pmidx,\pmidy)
  \global\advance\pbackx by  250
  \global\advance\pbacky by -250
  \put(\pbackx,\pbacky){$\bar{u}$}
  {\LARGE
     \global\advance\pbackx by 1000
     \global\advance\pbacky by 350
     \put(\pbackx,\pbacky){\}}
  }
  \global\advance\pbackx by 1250
  \global\advance\pbacky by  250
  \put(\pbackx,\pbacky){$K^-$}

  \Yfour=\Ysix
  \global\advance\Yfour by -1000
  \drawline\fermion[\E\REG](\Xfour,\Yfour)[\Xseven]
  \drawarrow[\E\ATTIP](\pmidx,\pmidy)
  \global\advance\pbackx by  250
  \global\advance\pbacky by -150
  \put(\pbackx,\pbacky){$u$}


  \Ysix=\photonbacky
  \global\advance\Ysix by -3000
  \put(\photonbackx,\Ysix){\oval(2000,2000)[l]}

  \Xfour=\photonbackx
  \Yfour=\Ysix
  \global\advance\Yfour by 1000
  \drawline\fermion[\E\REG](\Xfour,\Yfour)[\Xseven]
  \global\advance\pmidx by -800
  \drawarrow[\W\ATTIP](\pmidx,\pmidy)
  \global\advance\pbackx by  250
  \global\advance\pbacky by -400
  \put(\pbackx,\pbacky){$\bar{s}$}
  {\LARGE
     \global\advance\pbackx by 1000
     \global\advance\pbacky by 350
     \put(\pbackx,\pbacky){\}}
  }
  \global\advance\pbackx by 1250
  \global\advance\pbacky by  100
  \put(\pbackx,\pbacky){$K^+$}

  \Yfour=\Ysix
  \global\advance\Yfour by -1000
  \drawline\fermion[\E\REG](\Xfour,\Yfour)[\Xseven]
  \drawarrow[\E\ATTIP](\pmidx,\pmidy)
  \global\advance\pbackx by  250
  \global\advance\pbacky by -250
  \put(\pbackx,\pbacky){$s$}

  \put(10000,-2000){(c)}


  \Xone=26250
  \Yone=0
  \Xtwo=12000  
   
  \drawline\fermion[\E\REG](\Xone,\Yone)[\Xtwo]
 
  \Xthree=\pfrontx
  \Xfour=\Xtwo
  \global\divide\Xfour by 2
  \global\advance\Xthree by \Xfour
  \drawarrow[\W\ATTIP](\Xthree,\pfronty)
   
  \Xthree=\pfrontx
  \Xfour=\pbackx
  \global\advance\Xthree by -1000
  \global\advance\Xfour by 1000

  \global\advance\pfrontx by -1000
  \global\advance\pfronty by -250
  \put(\pfrontx,\pfronty){$\bar{u}$}
  {\LARGE
     \global\advance\pfrontx by -1250
     \global\advance\pfronty by  5000
  }
  \global\advance\pfrontx by -1500
  \put(\pfrontx,\pfronty){$D^0$}

  \global\advance\pbackx by  250
  \global\advance\pbacky by -250
  \put(\pbackx,\pbacky){$\bar{u}$}
  {\LARGE
     \global\advance\pbackx by 1000
     \global\advance\pbacky by 350
     \put(\pbackx,\pbacky){\}}
  }
  \global\advance\pbackx by 1250
  \global\advance\pbacky by  250
  \put(\pbackx,\pbacky){$\pi^-$}

  \global\advance\Yone by 10000
  \drawline\fermion[\E\REG](\Xone,\Yone)[\Xtwo]
     
  \Xthree=\pfrontx
  \Xfour=\Xtwo
  \global\divide\Xfour by 6
  \global\advance\Xthree by \Xfour  
  \drawarrow[\E\ATTIP](\Xthree,\pfronty)
   
  \Xthree=\pfrontx
  \Xfour=\Xtwo
  \global\divide\Xfour by 3
  \global\multiply\Xfour by 2
  \global\advance\Xthree by \Xfour  
  \drawarrow[\E\ATTIP](\Xthree,\pfronty)

  \Xthree=\pfrontx
  \Xfour=\pbackx
  \global\advance\Xthree by -1000
  \global\advance\Xfour by 1000

  \global\advance\pfrontx by -1000
  \global\advance\pfronty by -250
  \put(\pfrontx,\pfronty){$c$}

  \global\advance\pbackx by  250 
  \global\advance\pbacky by -250
  \put(\pbackx,\pbacky){$s$}
  

  \Xthree=\pfrontx
  \Xfour=\Xtwo
  \global\divide\Xfour by 3
  \global\advance\Xthree by \Xfour  
  \drawline\photon[\SE\REG](\Xthree,\pmidy)[8]
  \global\advance\Xthree by -300
  \Ythree=\Yone
  \global\advance\Ythree by -4000
  \put(\Xthree,\Ythree){$W^+$}
   


  \global\advance\photonbackx by 1000
  \put(\photonbackx,\photonbacky){\oval(2000,2000)[l]}

  \Xseven=\Xtwo
  \global\divide\Xseven by 4

  \Xfour=\photonbackx
  \Yfour=\photonbacky
  \global\advance\Yfour by 1000
  \drawline\fermion[\E\REG](\Xfour,\Yfour)[\Xseven]
  \global\advance\pmidx by -800
  \drawarrow[\W\ATTIP](\pmidx,\pmidy)
  \global\advance\pbackx by  250
  \global\advance\pbacky by -500
  \put(\pbackx,\pbacky){$\bar{d}$}
  {\LARGE
     \global\advance\pbackx by 1000
     \global\advance\pbacky by 500
     \put(\pbackx,\pbacky){\}}
  }
  \global\advance\pbackx by 1250
  \global\advance\pbacky by  150
  \put(\pbackx,\pbacky){$\pi^+$}

  \Yfour=\photonbacky
  \global\advance\Yfour by -1000
  \drawline\fermion[\E\REG](\Xfour,\Yfour)[\Xseven]
  \drawarrow[\E\ATTIP](\pmidx,\pmidy)
  \global\advance\pbackx by  250
  \global\advance\pbacky by -150
  \put(\pbackx,\pbacky){$u$}


  \Ysix=\photonbacky
  \global\advance\Ysix by 3000
  \put(\photonbackx,\Ysix){\oval(2000,2000)[l]}

  \Xfour=\photonbackx
  \Yfour=\Ysix
  \global\advance\Yfour by 1000
  \drawline\fermion[\E\REG](\Xfour,\Yfour)[\Xseven]
  \global\advance\pmidx by -800
  \drawarrow[\W\ATTIP](\pmidx,\pmidy)
  \global\advance\pbackx by  250
  \global\advance\pbacky by -250
  \put(\pbackx,\pbacky){$\bar{u}$}
  {\LARGE
     \global\advance\pbackx by 1000
     \global\advance\pbacky by 350
     \put(\pbackx,\pbacky){\}}
  }
  \global\advance\pbackx by 1000
  \global\advance\pbackx by  250
  \put(\pbackx,\pbacky){$K^-$}

  \Yfour=\Ysix
  \global\advance\Yfour by -1000
  \drawline\fermion[\E\REG](\Xfour,\Yfour)[\Xseven]
  \drawarrow[\E\ATTIP](\pmidx,\pmidy)
  \global\advance\pbackx by  250
  \global\advance\pbacky by -150
  \put(\pbackx,\pbacky){$u$}


  \Ysix=\photonbacky
  \global\advance\Ysix by -3000
  \put(\photonbackx,\Ysix){\oval(2000,2000)[l]}

  \Xfour=\photonbackx
  \Yfour=\Ysix
  \global\advance\Yfour by 1000
  \drawline\fermion[\E\REG](\Xfour,\Yfour)[\Xseven]
  \global\advance\pmidx by -800
  \drawarrow[\W\ATTIP](\pmidx,\pmidy)
  \global\advance\pbackx by  250
  \global\advance\pbacky by -400
  \put(\pbackx,\pbacky){$\bar{d}$}
  {\LARGE
     \global\advance\pbackx by 1000
     \global\advance\pbacky by 350
     \put(\pbackx,\pbacky){\}}
  }
  \global\advance\pbackx by 1250
  \global\advance\pbacky by  250
  \put(\pbackx,\pbacky){$\pi^+$}

  \Yfour=\Ysix
  \global\advance\Yfour by -1000
  \drawline\fermion[\E\REG](\Xfour,\Yfour)[\Xseven]
  \drawarrow[\E\ATTIP](\pmidx,\pmidy)
  \global\advance\pbackx by  250
  \global\advance\pbacky by -250
  \put(\pbackx,\pbacky){$d$}

  \put(31000,-2000){(d)}
\end{picture}

\vspace{1em}
\begin{picture}(43000,12000)

  \Xone=4000
  \Yone=0
  \Xtwo=12000  
  

  \drawline\fermion[\E\REG](\Xone,\Yone)[\Xtwo]

  \Xthree=\pfrontx
  \Xfour=\Xtwo
  \global\divide\Xfour by 4
  \global\advance\Xthree by \Xfour  
  \drawarrow[\E\ATTIP](\Xthree,\pfronty)
  
  \Xthree=\pfrontx
  \Xfour=\Xtwo
  \global\divide\Xfour by 4
  \global\multiply\Xfour by 3
  \global\advance\Xthree by \Xfour  
  \drawarrow[\E\ATTIP](\Xthree,\pfronty)
  
  \Xthree=\pfrontx
  \Ythree=\pfronty

  \Xfour=\pbackx
  \Yfour=\pbacky

  \global\advance\Xthree by -750
  \global\advance\Ythree by -250

  \global\advance\Xfour by  500
  \global\advance\Yfour by -250

  \put(\Xthree,\Ythree){$c$}
  \put(\Xfour,\Yfour){$s$}
  {\LARGE
     \global\advance\pbackx by 1500
     \global\advance\pbacky by 150
     \put(\pbackx,\pbacky){\}}
  }
  \global\advance\pbackx by 1250
  \global\advance\pbacky by 250
  \put(\pbackx,\pbacky){$\phi$}

  {\LARGE
     \global\advance\Xthree by -1250
     \global\advance\Ythree by  5000
   }
  \global\advance\Xthree by -1500
  \put(\Xthree,\Ythree){$D^0$}


  \Xthree=\pfrontx
  \Xfour=\Xtwo
  \global\divide\Xfour by 2
  \global\advance\Xthree by \Xfour
  \drawline\photon[\N\REG](\Xthree,\pfronty)[9]
  \global\advance\Xthree by -2500
  \Ythree=\photonfronty
  \global\advance\Ythree by 4000
  \put(\Xthree,\Ythree){$W^+$}
  

  \drawline\fermion[\E\REG](\Xone,\photonbacky)[\Xtwo]
  \Xthree=\pfrontx
  \Xfour=\Xtwo
  \global\divide\Xfour by 4
  \global\advance\Xthree by \Xfour
  \drawarrow[\W\ATTIP](\Xthree,\pfronty)
  \global\advance\Xthree by \Xfour
  \global\advance\Xthree by \Xfour
  \drawarrow[\W\ATTIP](\Xthree,\pbacky)
  
  \Xthree=\pfrontx
  \Ythree=\pfronty
  \global\advance\Xthree by -1000
  \global\advance\Ythree by -250
  \put(\Xthree,\Ythree){$\bar{u}$}

  \Xfour=\pbackx
  \Yfour=\pbacky
  \global\advance\Xfour  by 500
  \global\advance\Yfour by -250
  \put(\Xfour,\Yfour){$\bar{d}$}

  

  \Xsix=\photonbackx
  \Ysix=\photonbacky
  \global\advance\Xsix by  3000
  \global\advance\Ysix by -2500

  \put(\Xsix,\Ysix){\oval(2000,2500)[l]}

  \Xseven=\Xtwo
  \global\divide\Xseven by 4
  \global\advance\Ysix by 1225
  \drawline\fermion[\E\REG](\Xsix,\Ysix)[\Xseven]
  \global\advance\pmidx by 400
  \drawarrow[\E\ATTIP](\pmidx,\pmidy)
  \global\advance\pbackx by  500
  \global\advance\pbacky by -250
  \put(\pbackx,\pbacky){$u$}
  {\LARGE
     \global\advance\pbackx by 1000
     \global\advance\pbacky by 500
     \put(\pbackx,\pbacky){\}}
  }
  \global\advance\pbackx by 1250
  \global\advance\pbacky by  250
  \put(\pbackx,\pbacky){$\pi^+$}

  \Xseven=\Xtwo
  \global\divide\Xseven by 4
  \global\advance\Ysix by -2500
  \drawline\fermion[\E\REG](\Xsix,\Ysix)[\Xseven]
  \global\advance\pmidx by -400
  \drawarrow[\W\ATTIP](\pmidx,\pmidy)
  \global\advance\pbackx by  500
  \global\advance\pbacky by -250
  \put(\pbackx,\pbacky){$\bar{u}$}


  \Xsix=\photonfrontx
  \Ysix=\photonfronty
  \global\advance\Xsix by  3000
  \global\advance\Ysix by  2500

  \put(\Xsix,\Ysix){\oval(2000,2500)[l]}

  \Xseven=\Xtwo
  \global\divide\Xseven by 4
  \global\advance\Ysix by 1225
  \drawline\fermion[\E\REG](\Xsix,\Ysix)[\Xseven]
  \global\advance\pmidx by 400
  \drawarrow[\E\ATTIP](\pmidx,\pmidy)
  \global\advance\pbackx by  500
  \global\advance\pbacky by -250
  \put(\pbackx,\pbacky){$s$}
  {\LARGE
     \global\advance\pbackx by 1000
     \global\advance\pbacky by 500
     \put(\pbackx,\pbacky){\}}
  }
  \global\advance\pbackx by  750
  \global\advance\pbackx by  250
  \put(\pbackx,\pbacky){$K^-$}

  \Xseven=\Xtwo
  \global\divide\Xseven by 4
  \global\advance\Ysix by -2500
  \drawline\fermion[\E\REG](\Xsix,\Ysix)[\Xseven]
  \global\advance\pmidx by -400
  \drawarrow[\W\ATTIP](\pmidx,\pmidy)
  \global\advance\pbackx by  500
  \global\advance\pbacky by -250
  \put(\pbackx,\pbacky){$\bar{s}$}

  \put(10000,-3000){(e)}


  \Xone=26250
  \Yone=0
  \Xtwo=12000  
  

  \drawline\fermion[\E\REG](\Xone,\Yone)[\Xtwo]

  \Xthree=\pfrontx
  \Xfour=\Xtwo
  \global\divide\Xfour by 4
  \global\advance\Xthree by \Xfour  
  \drawarrow[\E\ATTIP](\Xthree,\pfronty)
  
  \Xthree=\pfrontx
  \Xfour=\Xtwo
  \global\divide\Xfour by 4
  \global\multiply\Xfour by 3
  \global\advance\Xthree by \Xfour  
  \drawarrow[\E\ATTIP](\Xthree,\pfronty)
  
  \Xthree=\pfrontx
  \Ythree=\pfronty

  \Xfour=\pbackx
  \Yfour=\pbacky

  \global\advance\Xthree by -750
  \global\advance\Ythree by -250

  \global\advance\Xfour by  500
  \global\advance\Yfour by -250

  \put(\Xthree,\Ythree){$c$}
  \put(\Xfour,\Yfour){$s$}
  {\LARGE
     \global\advance\pbackx by 1500
     \global\advance\pbacky by 150
     \put(\pbackx,\pbacky){\}}
  }
  \global\advance\pbackx by 1250
  \global\advance\pbacky by 250
  \put(\pbackx,\pbacky){$K^-$}

  {\LARGE
     \global\advance\Xthree by -1250
     \global\advance\Ythree by  5000
   }
  \global\advance\Xthree by -1500
  \put(\Xthree,\Ythree){$D^0$}


  \Xthree=\pfrontx
  \Xfour=\Xtwo
  \global\divide\Xfour by 2
  \global\advance\Xthree by \Xfour
  \drawline\photon[\N\REG](\Xthree,\pfronty)[9]
  \global\advance\Xthree by -2500
  \Ythree=\photonfronty
  \global\advance\Ythree by 4000
  \put(\Xthree,\Ythree){$W^+$}
  

  \drawline\fermion[\E\REG](\Xone,\photonbacky)[\Xtwo]
  \Xthree=\pfrontx
  \Xfour=\Xtwo
  \global\divide\Xfour by 4
  \global\advance\Xthree by \Xfour
  \drawarrow[\W\ATTIP](\Xthree,\pfronty)
  \global\advance\Xthree by \Xfour
  \global\advance\Xthree by \Xfour
  \drawarrow[\W\ATTIP](\Xthree,\pbacky)
  
  \Xthree=\pfrontx
  \Ythree=\pfronty
  \global\advance\Xthree by -1000
  \global\advance\Ythree by -250
  \put(\Xthree,\Ythree){$\bar{u}$}

  \Xfour=\pbackx
  \Yfour=\pbacky
  \global\advance\Xfour  by 500
  \global\advance\Yfour by -250
  \put(\Xfour,\Yfour){$\bar{d}$}

  

  \Xsix=\photonbackx
  \Ysix=\photonbacky
  \global\advance\Xsix by  3000
  \global\advance\Ysix by -2500

  \put(\Xsix,\Ysix){\oval(2000,2500)[l]}

  \Xseven=\Xtwo
  \global\divide\Xseven by 4
  \global\advance\Ysix by 1225
  \drawline\fermion[\E\REG](\Xsix,\Ysix)[\Xseven]
  \global\advance\pmidx by 400
  \drawarrow[\E\ATTIP](\pmidx,\pmidy)
  \global\advance\pbackx by  500
  \global\advance\pbacky by -250
  \put(\pbackx,\pbacky){$u$}
  {\LARGE
     \global\advance\pbackx by 1000
     \global\advance\pbacky by 500
     \put(\pbackx,\pbacky){\}}
  }
  \global\advance\pbackx by 1250
  \global\advance\pbacky by  250
  \put(\pbackx,\pbacky){$\pi^+$}

  \Xseven=\Xtwo
  \global\divide\Xseven by 4
  \global\advance\Ysix by -2500
  \drawline\fermion[\E\REG](\Xsix,\Ysix)[\Xseven]
  \global\advance\pmidx by -400
  \drawarrow[\W\ATTIP](\pmidx,\pmidy)
  \global\advance\pbackx by  500
  \global\advance\pbacky by -250
  \put(\pbackx,\pbacky){$\bar{u}$}


  \Xsix=\photonfrontx
  \Ysix=\photonfronty
  \global\advance\Xsix by  3000
  \global\advance\Ysix by  2500

  \put(\Xsix,\Ysix){\oval(2000,2500)[l]}

  \Xseven=\Xtwo
  \global\divide\Xseven by 4
  \global\advance\Ysix by 1225
  \drawline\fermion[\E\REG](\Xsix,\Ysix)[\Xseven]
  \global\advance\pmidx by 400
  \drawarrow[\E\ATTIP](\pmidx,\pmidy)
  \global\advance\pbackx by  500
  \global\advance\pbacky by -250
  \put(\pbackx,\pbacky){$u$}
  {\LARGE
     \global\advance\pbackx by 1000
     \global\advance\pbacky by 500
     \put(\pbackx,\pbacky){\}}
  }
  \global\advance\pbackx by 1250
  \global\advance\pbacky by  250
  \put(\pbackx,\pbacky){$\rho$}

  \Xseven=\Xtwo
  \global\divide\Xseven by 4
  \global\advance\Ysix by -2500
  \drawline\fermion[\E\REG](\Xsix,\Ysix)[\Xseven]
  \global\advance\pmidx by -400
  \drawarrow[\W\ATTIP](\pmidx,\pmidy)
  \global\advance\pbackx by  500
  \global\advance\pbacky by -250
  \put(\pbackx,\pbacky){$\bar{u}$}

  \put(31000,-3000){(f)}

\end{picture}

\vspace{5ex}
\renewcommand{\baselinestretch}{1.2}
\caption[Feynman diagrams for amlitudes that can contribute to the decays 
	  $ D^0 \to K^- K^- K^+ \pi^+ $ and
          $ D^0 \to K^- \pi^- \pi^+ \pi^+ $.]
        {Feynman diagrams for amplitudes that can contribute to the decays 
	  $ D^0 \to K^- K^- K^+ \pi^+ $ and 
          $ D^0 \to K^- \pi^- \pi^+ \pi^+ $.
	 When one light quark pair (either $d \bar{d}$ or $u \bar{u}$)
	 in the latter is replaced by an $s \bar{s}$
	 pair, we get the former.}
\label{fDiagrams}
\end{center}
\end{figure}

Neither decay is purely non-resonant, but we can use 
the value of $ {\cal R}, $ determined using this equation, to estimate crudely
the importance of amplitudes in which at least one $ q \overline q $ pair is produced
from the vacuum.
The decay $ D^0 \to K^- \pi^- \pi^+ \pi^+ $  can proceed via amplitudes
in which the quarks produced in a spectator decay coalesce to form 
hadrons which, in turn,  decay strongly to produce four hadrons in the final
state.
The decay $ D^0 \to K^- K^- K^+ \pi^+ $ cannot proceed via such amplitudes.
It requires either that an extra $ s \overline s $ pair be produced from
the vacuum or that long distance final state interactions of hadrons
produced at short distances produce such pairs.
Feynman diagrams for such amplitudes are shown in Fig.\  \ref{fDiagrams},
along with those of corresponding amplitudes for $ D^0 \to K^- \pi^-
\pi^+ \pi^+ $ decays where an extra $ u \overline u $ or $ d \overline d $
pair is produced.

Accounting for the differences in phase spaces using 
$\Omega_{K K K \pi} / \Omega_{K \pi \pi \pi} $, and ignoring
any possible quantum mechanical intereferences,
one can write $ \cal R $ in terms of the probabilities that the
final states we are considering are produced by amplitudes
in which a $ q \overline q $ state is produced from the
vacuum,  $ P_{u \overline u} $,
$ P_{d \overline d} $, $ P_{s \overline s} $ for $ u \overline u $,
$ d \overline d $, and $ s \overline s $ respectively, or in which
the amplitude has no pair produced from the vacuum,
$ P_{\rm no \ pair} $.
As a first approximation, one can imagine
a form of isospin symmetry in which  
$  P_{d \overline d} =
P_{s \overline s} $ for each amplitude that can lead to four
charged hadrons in the final state and the
corresponding $ u \overline u $
amplitude does not lead to
four charged hadrons in the final state.
In this case we calculate
\begin{equation}
{\cal R} = 
0.32 = 
{    
 { P_{s \overline s}} 
\over
 {P_{s \overline s} + P_{\rm no \ pair} }
}
\end{equation}
in which case 
\begin{equation}
{  P_{s \overline s} \over P_{\rm no \ pair}  }= 0.47 \ .
\end{equation}
If the amplitudes for producing each flavor of
$ q \overline q $ pair are the same, and the likelihoods
for producing four charged hadrons in the final state are the same
(allowing for resonant three-body decays as well as for
non-resonant four-body decay),
then one might expect
$ P_{u \overline u} $ = $ P_{d \overline d} $ = 
$ P_{s \overline s } $.
In this case 
\begin{equation}
{  P_{s \overline s} \over P_{\rm no \ pair}  }= 0.90 \ .
\end{equation}
If the amplitudes with $ u \overline u $ and $ d \overline d $
pairs produced from the vacuum somehow interfere 
destructively so that the
$ D^0 \to K^- \pi^- \pi^+ \pi^+ $ decay rate is equal to
that which would be produced in the absence of these
additional amplitudes, $ {\cal R} $ is a direct measurement of
\begin{equation}
{  P_{s \overline s} \over P_{\rm no \ pair}  } = 0.32 \ .
\end{equation}
A simple measurement of
$ \Gamma ( D^0 \to K^- K^- K^+ \pi^+ ) /
  \Gamma ( D^0 \to K^- \pi^- \pi^+ \pi^+ ) $
cannot tell us which picture is closest to the truth,
although it seems likely that
$ 0.3 < {  P_{s \overline s} / P_{\rm no \ pair}  }
<0.9 $.

\medskip

\bigskip

\noindent{\bf Summary}

Using data from Fermilab experiment E791, 
we have studied the decay $ D^0 \to K^- K^- K^+ \pi^+ $.
To avoid bias,
the selection criteria for the $  K^- K^- K^+ \pi^+ $
candidates were determined ``blindly" --
we masked the signal region
in the real data and systematically studied sensitivity using a 
combination of real data for background, and Monte Carlo simulations
and real data in the normalization decay mode 
$ D^0 \to K^- \pi^- \pi^+ \pi^+ $ for signal.
Only after we had determined the final set of cuts
did we examine
the data in the signal region.
We observe a signal of $ 18.4 \pm 5.3 $ events
from which we find the ratio of decay rates
$ \Gamma ( D^0 \to K^- K^- K^+ \pi^+ ) /
  \Gamma ( D^0 \to K^- \pi^- \pi^+ \pi^+ ) $
to be $ ( 0.54 \pm 0.16 \pm 0.08 ) \% $.
We also have examined the $ K^- K^+ $ invariant mass distribution
of signal events looking for evidence of resonant substructure,
{\em i.e.}, $ D^0 \to \phi K^- \pi^+$; 
$\phi \to K^- K^+ $.
Fitting the distribution using an incoherent sum of resonant
and non-resonant signal shapes plus a background shape, we
find that $ 0.7 \pm 0.3 $ of the signal comes from
$ \phi K \pi $.
Finally, using the ratio of non-resonant phase-spaces for the
two decays as an approximation for the correctly weighted
ratio, we find the ratio of matrix elements that lead to
the signal and normalization final states to be
$ {\cal R} = 0.32 \pm 0.10 $.
Producing $ D^0 \to K^- K^- K^+ \pi^+ $ requires producing
an extra $ s \overline s $ pair from the vacuum or in
a final state interaction. 
Relating this probability to $ {\cal R} $ is highly
model-dependent, and our measurement does not suffice to
distinguish among models.
However, it seems likely that 
$ 0.3 \, < \,  P_{s \overline s} / P_{\rm no \ pair} 
\, < \, 0.9$.

\bigskip

\noindent {\bf Acknowledgements}
 

	We gratefully acknowledge the assistance of the staffs of 
Fermilab and of all the participating institutions. This research was 
supported by the Brazilian Conselho Nacional de Desenvolvimento Cient\'{\i}fico 
e Tecnol\'ogico,
CONACYT (Mexico), the U.S.-Israel Binational Science Foundation, and the U.S.
National Science Foundation. Fermilab is operated by the Universities 
Research Associates, Inc., under  contract with the United States Department
of Energy.




\begin{thebibliography}{90}

\bibitem{e791} 
E791 Collaboration, E.~M.~Aitala {\em et al.}, Eur.~Phys.~J.direct
{\bf C4}, 1 (1997);
J.~A.~Appel, Ann. Rev. Nucl. Part. Sci. {\bf 42}, 367 (1992), and 
references therein; 
D.~J. Summers {\em et al.}, in: Proceedings of the 
{\em XXVII${}^{\hbox {\rm \, th}}$ Rencontre de Moriond}, Electroweak Interactions
and Unified Theories, Les Arc, France, 417 (1992).

\bibitem{cerenkov} D. Bartlett {\em et al.}, Nucl. Instr. and Meth. {\bf A260},
55 (1986).

\bibitem{daq}  S. Amato {\it et al.}, Nucl.\ Instr.\ and Meth. {\bf A324},
535 (1993).

\bibitem{farm} F. Rinaldo and S. Wolbers,  Computers in Physics  {\bf  7},
184 (1993); S. Bracker {\it et al.},  IEEE Trans. Nucl. Sci. {\bf 43},
2457 (1996). 

\bibitem{e791_previous} E791 Collaboration, E.~M.~Aitala {\em et al.},
Phys. Lett {\bf B462}, 401 (1999).

\bibitem{e687} E687 Collaboration,  L. Frabetti    {\it et al.},  Phys. Lett.
{\bf B354},  486 (1995).
  

\bibitem{pdg} Review of Particle Physics,  R.~M.~Barnett {\it et al.},  Phys. 
Rev. D
{\bf 54}, 455 (1996).

\bibitem{lalith} E791 Collaboration, E. Aitala {\em et al.}, Phys. Lett.
{\bf B423}, 185 (1998).

\bibitem{markIII} MARK III Collaboration, D.~Coffman {\em et al.},
Phys. Rev. D {\bf 45}, 2196 (1992).


\end{thebibliography}
\end{document}